\documentclass[aps,pre,reprint,groupedaddress,showpacs]{revtex4-1}
\usepackage{amsmath,amssymb,amsxtra,amsfonts,graphicx,wrapfig}
\usepackage{xcolor}
\usepackage{subfigure}
\usepackage{enumerate}
\usepackage[font=scriptsize,labelfont=bf]{caption}
\usepackage{color}
\begin{document}
\title{Exclusion Process on two intersected lanes with constrained resources: Symmetry breaking and shock dynamics}
\author{Akriti Jindal}
\author{Arvind Kumar Gupta}
\email[]{akgupta@iitrpr.ac.in}
\affiliation{Department of Mathematics, Indian Institute of Technology Ropar, Rupnagar-140001, Punjab, India.}

\begin{abstract}
We present a study of exclusion process on a peculiar topology of network with two intersected lanes, competing for the particles in a reservoir with finite capacity. To provide a theoretical ground for our findings, we exploit mean-field approximation along with domain-wall theory. The stationary properties of the system including phase transitions, density profiles and position of the domain-wall are derived analytically. Under the similar dynamical rules, the particles of both the lanes interact only at the intersected site. The symmetry of system is maintained till number of particles do not exceed total number of sites. However, beyond this the symmetry breaking phenomenon occurs resulting in the appearance of asymmetric phases and continues to persist even for infinite number of particles. The complexity of phase diagram shows a non-monotonic behaviour with increasing number of particles in the system. A bulk induced shock appears in a symmetric phase, whereas, a boundary induced shock is observed in symmetric as well as asymmetric phase. Monitoring the location of shock with increasing entry of particles, we explain the possible phase transitions. The theoretical results are supported by extensive Monte Carlo simulations and explained using simple physical arguments.
\end{abstract}
\maketitle
\section{Introduction}
The decisive requirement for the functioning of any complex system ranging from the subcellular level of biological organisms
to globe-spanning man-made structures is the transportation of matter and information. From theoretical point of view, these stochastic transport phenomenon are an intriguing example of multi-particle systems addressing far from equilibrium processes. The extensive organisation of interconnected linelike pathways for transport mechanisms forms a network like structure. However, the study of complex frameworks still remains a major challenge in the field of physics and cellular biology. In this direction, the investigation of relatively simpler topologies is a crucial step towards understanding the complex network systems. 

In statistical physics, lattice gas exclusion processes have gained much popularity to model the active stochastic motion of particles along a one-dimensional lane \cite{schutz2000exactly,spitzer1991interaction,spohn2012large}. Specifically, totally asymmetric simple exclusion process (TASEP) has been a paradigmatic model to study the motion of self-propelled particles in one preferred direction subjected to excluded volume interactions \cite{derrida1992exact,parmeggiani2003phase}. It was originally introduced in the context of RNA polymerization by ribosomes \cite{schutz2000exactly,macdonald1968kinetics,macdonald1969concerning}. Since then these models have further stimulated a lot of fundamental research including vehicular traffic, intracellular transport, surface growth, transport in ion channels, etc \cite{macdonald1968kinetics,mallick2011some,schadschneider2000statistical,helbing2001traffic,chou2011non}.  In terms of TASEP, numerous simpler topologies like junctions, tree like structures, structureless links have been extensively analysed  in context of traffic flow and biological transportation \cite{neri2011totally,embley2009understanding,basu2010asymmetric,neri2013exclusion}.
However, the collective behaviour of particles on elementary structures of networks is still a subject of comprehensive discussion.

In the perspective of modelling generic features of transportation processes on lane-based systems, studies abound in the literature analysing the topology of intersecting lanes \cite{fouladvand2004optimized,thunig2016braess,crociani2016multiscale}.
Owing to an extensive body of TASEP models, quantitative characterisation of two crossing roads with parallel update rules in a closed geometry has been well examined \cite{ishibashi1996phase,ishibashi2001phase,ishibashi2001phase1}. Recently, a variant topology considering ``figure-of-eight" network of two intersecting TASEPs has been studied with periodic conditions following random sequential update rules \cite{braess2005paradox}. This prototype provides an insight to the Braess paradox that explains counter-intuitive situation in which
adding an edge to a road network leads to a user optimum with higher travel times for all network users \cite{pas1997braess}. However, the dynamics of two intersecting lanes with open boundary conditions revealed a much interesting phenomenon known as spontaneous symmetry breaking (SSB) \cite{yuan2008spontaneous,tian2020totally}. This displays the occurrence of macroscopic asymmetric stationary states under the symmetric microscopic dynamical rules. The``bridge model" was the pioneer model to exhibit SSB where two species of particles are allowed to move in opposite direction on a single lane TASEP \cite{evans1995spontaneous,evans1995asymmetric}. Since then this aspect has been of specific interest studied in detail utilising variants of TASEP assimilating various additional processes \cite{muhuri2011phase,popkov2008spontaneous,pronina2007spontaneous,sharma2017phase,verma2018far}. Such models have been thoroughly investigated in the vicinity of unlimited supply of particles.

In recent years, much more generalised versions of TASEP model have been contemplated where the particles are injected from a finite reservoir of particles. Such models have shown a wide applicability in many physical and biological systems such as protein synthesis, movement of motor proteins, ``parking garage problems", vehicular traffic \cite{adams2008far,ha2002macroscopic,blasius2013recycling}. To explore such scenario few studies based on single as well as multi-lane TASEP prototypes have been conducted where the dynamics compete for limited number of particles \cite{verma2018limited,greulich2012mixed,adams2008far,cook2009feedback,cook2009competition}. This generalisation reveals a non-trivial behavior of system including the extension of ``shock phase" that leads to traffic jam like situations on lanes. For this phase domain-wall approach provides a powerful theoretical technique to incorporate fluctuations and accurately calculate the stationary properties of the system \cite{cook2009competition}.
\begin{figure*}
\centering

\includegraphics[width = 0.43\textwidth]{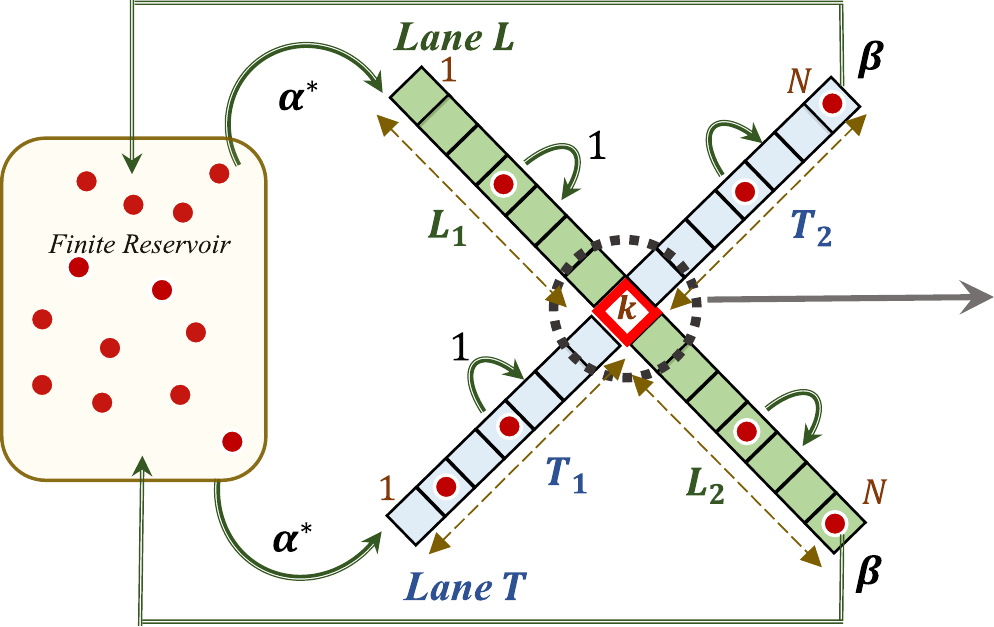}
\includegraphics[width = 0.24\textwidth]{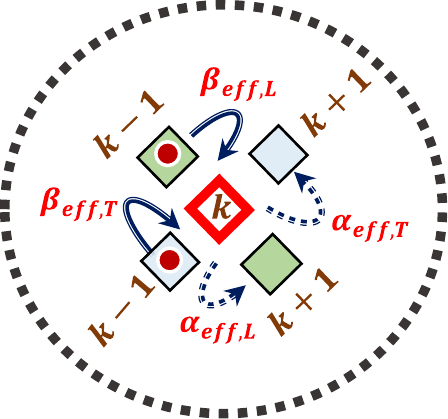}
\caption{\label{fig1}Schematic representation of two intersected lanes in a reservoir with finite number of particles. Green and blue coloured lane represents lane $L$ and $T$ respectively with $i=1,2,\hdots,N$ labelled sites. The intersected site $k=N/2$ is highlighted in red colour that can occupy any of the particle arriving from lane $L$ or $T$. If this site accommodates particle from $L(T)$ it can jump to the unoccupied site $k+1$ of same lane with unit rate. The entry and exit rate of particles in either lane is $\alpha^*$ and $\beta$. Lane $L(T)$ is divided into two parts left segment $L_1(T_1)$ and right segment $L_2(T_2)$. The particles can leave $L_1(T_1)$ with effective exit rate $\beta_{eff,L}(\beta_{eff,T})$ and enter $L_2(T_2)$ with effective entry rate $\alpha_{eff,L}(\alpha_{eff,T})$.}
\end{figure*}

In this work, we present a unifying picture for the emergence of shock and symmetry breaking on two-intersecting lanes coupled to a reservoir of finite number of particles. We attempt to provide an insight to mechanisms including transportation processes where the particles can conquer a jam like situation. We treat this model as a two-lane coupled system with an inhomogeneity for which we exploit the idea of effective rates and domain wall theory to analyse the interplay of intersected site and finite reservoir. We attempt to answer few important questions like (1) Does symmetry breaking prevail in a system of intersected lanes under the effect of finite resources? (2) Are there non-trivial effects on the qualitative and quantitative behaviour on phase diagram with variation of number of particles in the system? We also provide a fundamental brief by considering appropriate limiting cases to visualise the steady-state characteristics of the system.

\section{Model Definition and Dynamical Rules}
This section intends to elaborate a minimalistic model of two-lane transport intersecting at a special site. The extreme ends of both the lanes are coupled to a single reservoir having finite number of identical particles denoted by $N_r$. The total number of particles $(N_{tot})$ in the system remain constant at any instant of time. The two lanes are labelled as $L$ and $T$ assuming each lane to be composed of $i=1,2,\hdots,N$ sites with special site at $k=N/2$ common to both the lanes as shown in Fig. \eqref{fig1}. We assume this site far away from boundaries to probe the effect of intersected lanes on the overall dynamics of the system. The transition rules are in accordance to random sequential update rules. Each site including intersected site obeys hard core exclusion principle that allows each site to occupy atmost one particle. \par 
We presume particles in both the lanes to move in one preferred direction from left to right.
A particle is allowed to enter the vacant site $1$ of any of the two lanes from the reservoir with effective intrinsic rate $\alpha^{*}$ depending on the reservoir density given by,
\begin{equation}
\alpha^{*}=\alpha f(N_r)
\end{equation}
where $\alpha$ is entry rate for the case with infinite number of particles.

It is reasonable to adopt a monotonic increasing function satisfying $f(0)=0$ and $f(N_{tot})=1$. This means smaller the number of particles in the reservoir lower the effective intrinsic rate of particles in the two lanes. And, the enhanced particle content in the reservoir leads to greater rush of particles in the two lanes. Based on these arguments, the simplified choice for $f(N_r)$ is,
\begin{equation}
f(N_r)=\dfrac{N_r}{N_{tot}}
\end{equation}   
that implies the effective intrinsic rate given by,
\begin{equation}\label{alphaeff}
\alpha^{*}=\alpha \dfrac{N_r}{N_{tot}}.
\end{equation}
This relation implies that the entry rate of particles is directly proportional to the free concentration of particles in the reservoir as long as it is not too crowded. The choice of function is generic and suits well to imitate biological as well as vehicular transport processes \cite{adams2008far,ha2002macroscopic,blasius2013recycling}. For either lane, the exit rate of particles is independent of the number of particles present in the reservoir. A particle at site $N$ can escape with constant rate $\beta$ back to the reservoir from where it is free to rejoin any lane. 

In the bulk of each lane, a particle seeks to jump to the adjacent vacant site of the same lane with unit rate and are not allowed to switch their lanes. However, the intersection of two lanes at site $k$ distinguishes it from a homogeneous two-lane TASEP model. Since, site $k$ is shared by both the lanes, any of the particle approaching from lane $L$ or $T$ can occupy this site at any instant of time. A particle at site $k-1$ of lane $L(T)$ can jump to intersected site $k$ with unit rate if found empty. Further, if site $k$ is occupied with the particle arriving from lane $L$ or $T$, it is allowed to jump to the unoccupied site $k+1$ of its own respective lane with rate 1. Here, the particles are not allowed to change their lanes even after jumping from the intersected site. For the proposed model, particles of both the lanes interact only at intersected site because a particle at site $k-1$ of lane $L(T)$ compete to find an empty site $k$.

It has been noticed that for the case of infinite number of particles, the considered topology of lanes induces a non-trivial effect on the qualitative behaviour of system in terms of symmetry breaking \cite{tian2020totally}. However, in the previous work, the existence of phase regimes specially for asymmetric phase is calculated numerically.

\section{Theoretical Framework}
The present model can be viewed as a variant of TASEP model with two incoming segments reaching a junction site and diverging into two outgoing segments \cite{embley2009understanding}. This represents a network of $2\times 2$ segments provided the particles at site $k$ are distinguishable and are constrained to jump to the next site of specific lane. When these lanes do not intersect and there are infinite number of particles, the proposed model reduces to a two lane homogeneous TASEP model.

\subsection{Brief discussion of results for two-lane homogeneous TASEP model with infinite reservoir}
In literature, mean-field approximation has been discerned to provide exact results for homogeneous single-lane TASEP model with infinite reservoir \cite{derrida1992exact,chou2011non,kolomeisky1998phase}. It has been found that there exists three distinct stationary phases: low density (LD), high density (HD) and maximal current (MC) \cite{derrida1992exact,chou2011non,kolomeisky1998phase}. Various network TASEP models have been extensively explored utilising mean-field theory \cite{pronina2005theoretical,neri2013exclusion,neri2011totally,embley2009understanding,basu2010asymmetric,braess2005paradox}. This approximation ignores all possible correlation in the system and assumes the occupancy of two consecutive sites independent of each other. 

Based on this approach, it has been deliberated that for equal entry and exit rates of two uncoupled lanes ($L$ and $T$), there exist three dynamic regimes namely LD:LD, HD:HD and MC:MC \cite{derrida1992exact}. The bulk density in each phase remains equal for both the lanes, leading to the existence of symmetric phases. In the notation, first part of `:' denotes the state of lane $L$, and second part describes the phase manifested by lane $T$. The description of theoretically calculated density profiles, particle currents and existence of phases is summarised in table \eqref{simple}.

In the continuum limit, steady-state particle current in the bulk ($1<i<N$) of lane $L$ and $T$ denoted by $J_L$ and $J_T$ respectively, is written as,
\begin{equation}
J_L=\rho(1-\rho),\qquad J_T=\sigma(1-\sigma)
\end{equation}
where $\rho$ and $\sigma$ denotes the average bulk density in lane $L$ and $T$ respectively.
And, the current at site $1$ and $N$ of each lane is given by,
\begin{eqnarray}
J_L^1=\alpha(1-\rho^1),\qquad J_L^N=\rho^N\beta,\\
J_T^1=\alpha(1-\sigma^1),\qquad J_T^N=\sigma^N\beta
\end{eqnarray}
where $\rho^1(\sigma^1)$ and $\rho^N(\sigma^N)$ represents the average density of particles at site $1$ and $N$ of lane $L(T)$ respectively.

\begin{table*}[!htb]
\caption{\label{simple}Summary of results for a two uncoupled lanes TASEP model with finite reservoir \cite{greulich2012mixed}. When $\mu\rightarrow\infty$, $\alpha^*\rightarrow\alpha$, the results converge to that for model with infinite reservoir of particles \cite{derrida1992exact}. Here, $x_w$ denotes the position of domain wall.}
\begin{center}
\renewcommand{\arraystretch}{2}
\resizebox{0.9\textwidth}{!}{
\begin{tabular}{||c|c||c|c|c|c|c||}
\hline \hline
&Phase Region&$\rho^1=\sigma^1$&$\rho=\sigma$&$\rho^N=\sigma^N$&Current($J_L=J_T$)&$\alpha^*$\\
\hline \hline
LD:LD&$\alpha^*<\min\{\beta,1/2\}$&$\alpha^*$&$\alpha^*$&$\dfrac{\alpha^*(1-\alpha^*)}{\beta}$&$\alpha^*(1-\alpha^*)$&$\alpha\left(1-\dfrac{1}{\mu}\right)$\\
\hline
HD:HD&$\beta<\min\{\alpha^*,1/2\}$&$1-\dfrac{\beta(1-\beta)}{\alpha^*}$&$1-\beta$&$1-\beta$&$\beta(1-\beta)$&$\alpha\left(1-\dfrac{2(1-\beta)}{\mu}\right)$\\
\hline 
MC:MC&$1/2<\min\{\alpha^*,\beta\}$&$1-\dfrac{1}{4\alpha^*}$&$\dfrac{1}{2}$&$\dfrac{1}{4\beta}$&$\dfrac{1}{4}$&$\alpha\left(1-\dfrac{1}{\mu}\right)$\\
\hline
S:S&$\alpha^*=\beta,~\beta<1/2$&$\alpha^*$&$x_w\alpha^*+(1-\beta)(1-x_w)$&$1-\beta$&$\alpha^*(1-\alpha^*)=\beta(1-\beta)$&$\dfrac{1}{x_w}\left(\dfrac{\mu}{2\alpha}(\alpha-\beta)-(1-\beta)(1-x_w)\right)$\\
\hline \hline
\end{tabular}
}
\end{center}
\end{table*}
\subsection{Bulk dynamics: Intersection of two lanes}\label{intersectedsection}
The intersection of two lanes at a special site introduces an inhomogeneity in the system. For this, we divide each lane into two segments, left segment $L_1$, $T_1$: $i=1,2,\hdots,k-1$ and right segment $L_2$, $T_2$: $i=k+1,k+2,\hdots,N$ coupled at special site $k$. 
The two segments of both the lanes are properly integrated by determining the effective rate of particles. We define, particles in lane $L$ and $T$ can leave their respective left segment with effective exit rate $\beta_{eff,L}$ and $\beta_{eff,T}$ respectively. Similarly, the particles can enter the right segment of lane $L$ and $T$ with effective entry rate $\alpha_{eff,L}$ and $\alpha_{eff,T}$  respectively. The average density of particles at site $k-1$, $k$ and $k+1$ of lane $L$ is written as $\rho_1^{k-1}$, $\rho^{k}$ and $\rho_2^{k+1}$. Similarly, for lane $T$, to represent these densities $\rho$ is replaced by $\sigma$.
 
We denote the particle current induced in each lane as $J_{L_j}$ and $J_{T_j}$ where $j=1$ (left segment) and $j=2$ (right segment).
The stationary current arguments in both the lanes leads to equal current in two segments of each lane, read as,
\begin{equation}\label{current}
J_{L_1}=J_{L_2}\qquad \text{and} \qquad J_{T_1}=J_{T_2}
\end{equation}
that implies
\begin{equation}\label{eqcur}
\rho_1^b(1-\rho_1^b)=\rho_2^b(1-\rho_2^b),\quad \text{and} \quad\sigma_1^b(1-\sigma_1^b)=\sigma_2^b(1-\sigma_2^b)
\end{equation}
where $\rho_j^b$ and $\sigma_j^b$ denotes the bulk density in two segments of lane $L$ and $T$ respectively.
The above equations in Eqn. \eqref{eqcur} further specifies that the bulk density satisfies,
\begin{eqnarray}
\rho_1^b=\rho_2^b,\quad \text{or} \quad \rho_1^b+\rho_2^b=1,\\
\sigma_1^b=\sigma_2^b,\quad \text{or} \quad \sigma_1^b+\sigma_2^b=1.
\end{eqnarray}
In each lane, the condition of current continuity suggests that the exit current of left segment is equal to the current passing from site $k-1$ to $k$, given by,
\begin{eqnarray}
\rho_1^{k-1}~\beta_{eff,L}&=&\rho_1^{k-1}~(1-\rho^{k}-\sigma^{k}),\\
\sigma_1^{k-1}~\beta_{eff,T}&=&\sigma_1^{k-1}~(1-\rho^{k}-\sigma^{k}).
\end{eqnarray}
that results in,
\begin{equation}\label{ebeta}
\beta_{eff,L}=\beta_{eff,T}=1-\rho^{k}-\sigma^{k}=\beta_{eff}~(\text{say}).
\end{equation}
Also, the current entering into the right segment is equal to current passing from site $k$ to $k+1$, written as,  
\begin{eqnarray}
(1-\rho_2^{k+1})~\alpha_{eff,L}&=&\rho^{k} ~(1-\rho_2^{k+1}),\\
(1-\sigma_2^{k+1})~\alpha_{eff,T}&=&\sigma^{k} ~(1-\sigma_2^{k+1})
\end{eqnarray}
that leads to,
\begin{eqnarray}\label{inter}
\alpha_{eff,L}=\rho^{k},\\
\alpha_{eff,T}=\sigma^{k}.
\end{eqnarray}
Our main aim is to calculate the effective rates and average density of particles in $L$ and $T$ including at intersected site. The explicitly computed effective rates helps to determine the stationary properties of the system. When there are infinite number of particles in the system, it has been observed that the intersection of lanes assists the phenomenon of symmetry breaking. The appearance of two symmetric and one asymmetric phase has been reported \cite{tian2020totally}.

\subsection{Boundary dynamics: Lanes connected to finite reservoir of particles}
The extreme ends of two intersected lanes are coupled to a finite reservoir of particles that governs the effective intrinsic rate of particles into the lanes as given in Eqn. \eqref{alphaeff}. Now, the total number of particles in the system can be written as,
\begin{equation}
N_{tot}=N_r+N_L+N_T
\end{equation} 
where $N_L$ and $N_T$ signifies the number of particles in lane $L$ and $T$ respectively.

Since, each lane is divided in two segments, the average density of particles in each lane is the sum of average density of particles in the respective left and right segment. Since, in the continuum limit, the spatial variables are rescaled as $x=i/N$, therefore, we can write the average density of particles in lane $L$ and $T$ as,
\begin{eqnarray}
\int_0^{1}\rho ~dx &=&\int_0^{1/2}\rho_1^b ~dx+\int_{1/2}^{1}\rho_2^b~ dx,\\
\int_0^{1}\sigma ~dx &=&\int_0^{1/2}\sigma_1^b ~dx+\int_{1/2}^{1}\sigma_2^b ~dx
\end{eqnarray}
respectively, where $\int_0^{1}\rho~dx=\dfrac{N_L}{N}$ and $\int_0^{1}\sigma~dx=\dfrac{N_T}{N}$.
Further, the effective intrinsic rate in Eqn. \eqref{alphaeff} can be written as,
\begin{eqnarray}
\alpha^* &=&\alpha-\dfrac{\alpha N}{N_{tot}}\left(\int_0^{1/2}(\rho_1^b+\sigma_1^b)~ dx+\int_{1/2}^{1}(\rho_2^b+\sigma_2^b)~ dx\right),\qquad
\end{eqnarray}
that implies
\begin{eqnarray}\label{finite}
\alpha^* &=&\alpha-\dfrac{\alpha}{\mu}\left(\int_0^{1/2}(\rho_1^b+\sigma_1^b)~ dx+\int_{1/2}^{1}(\rho_2^b+\sigma_2^b)~ dx\right),\qquad
\end{eqnarray}
where $\mu=\dfrac{N_{tot}}{N}$. As already observed in table \eqref{simple} that the density of particles is distinct in each phase, therefore the effective intrinsic rate $\alpha^*$ alters correspondingly. 
For the case when two lanes do not intersect, we retrieve a two-lane homogeneous model coupled to a finite reservoir of particles. Therefore, the expression for effective intrinsic rate in Eqn. \eqref{finite} reduces to \cite{verma2018limited},
\begin{equation}
\alpha^* =\alpha\left(1-\dfrac{1}{\mu}\int_0^{1}(\rho+\sigma)~ dx
\right).
\end{equation}  
Here, in addition to existing distinct stationary phases for homogeneous two-lane model with infinite reservoir, a new symmetric coexistence LD-HD phase namely shock (S) phase has been observed \cite{greulich2012mixed}.
A shock phase is characterised by an existing discontinuity in the bulk connected by a low-density to high-density regime. In literature, a well-known approach `` domain-wall theory" has been deployed to estimate the position of this domain-wall \cite{de1989microscopic,kolomeisky1998phase,verma2018limited}. The basic idea of this theory is to assume a sharp shock in the density profile located at any site between the two regimes. The estimation of locating this shock at a particular site helps to evaluate the overall density profile of particles. This shock is situated anywhere between [0,1] and its position is denoted by $x_w$ throughout the paper. The conditions for the existence of stationary phases with finite reservoir have been reviewed in table \eqref{simple}.

In the next section, we discuss the conditions for the existence of different phase regimes in ($\alpha,\beta$) plane for the proposed inhomogeneous model of intersecting lanes. The phase boundaries are obtained theoretically utilising the framework adopted in section \eqref{intersectedsection} along with the concept of domain-wall theory. In addition, we present the convergence of our theoretical results to the limiting case of intersected lanes with infinite reservoir \cite{tian2020totally}.
\begin{figure*}[!htb]
\subfigure[\label{001}$\mu=0.001$]{\includegraphics[width = 0.3\textwidth]{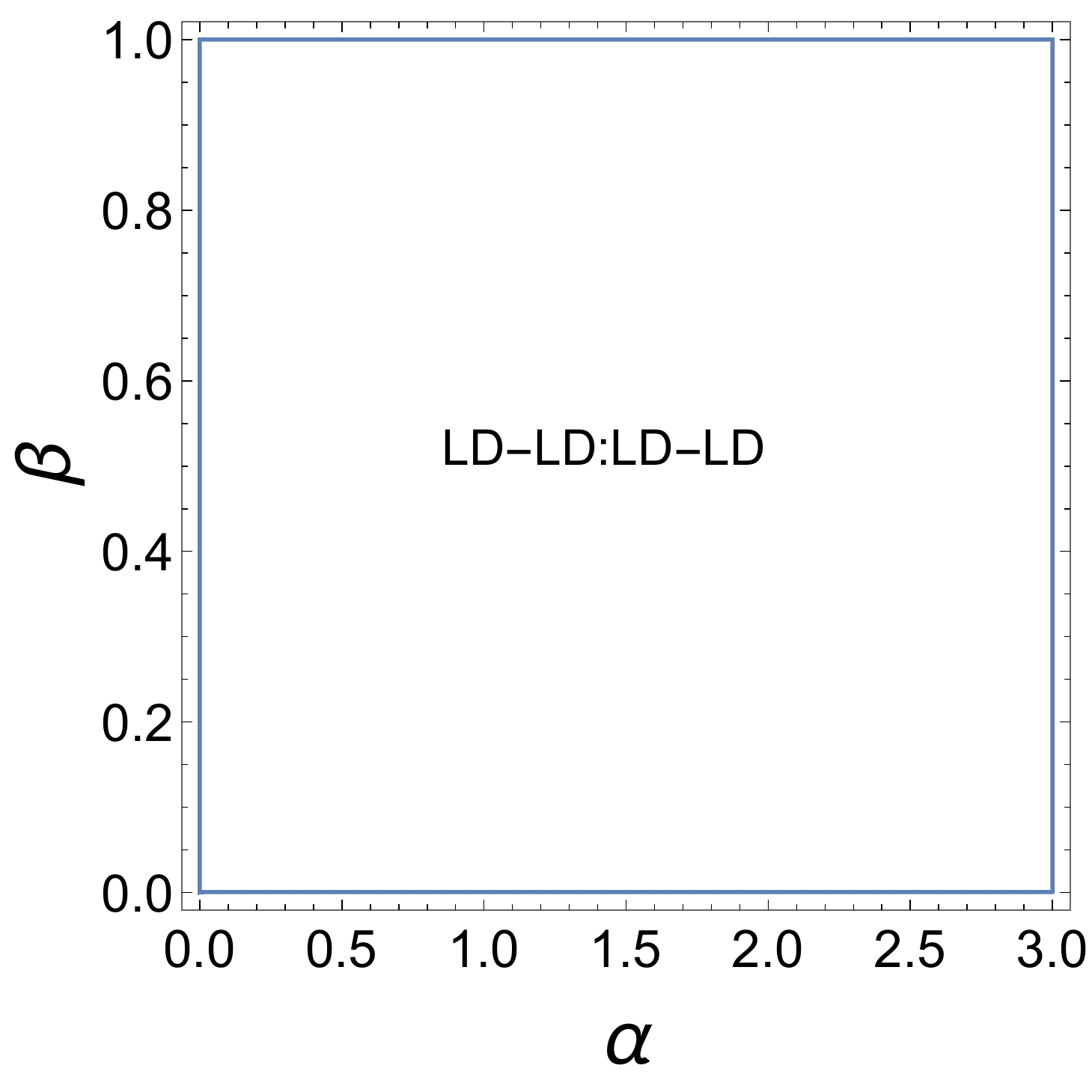}}
\subfigure[\label{05}$\mu=0.5$]{\includegraphics[width = 0.3\textwidth]{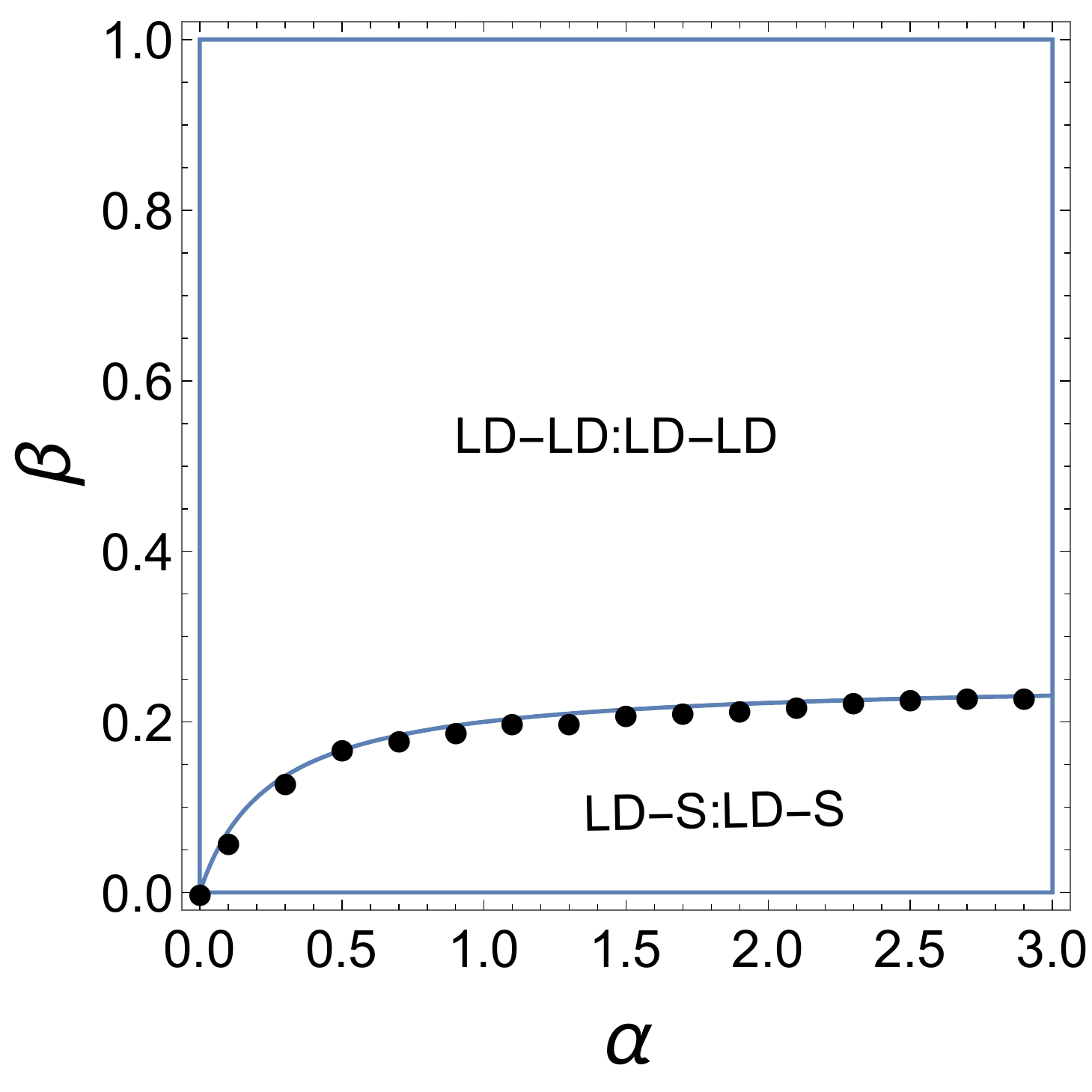}}
\subfigure[\label{1}$\mu=1$]{\includegraphics[width = 0.3\textwidth]{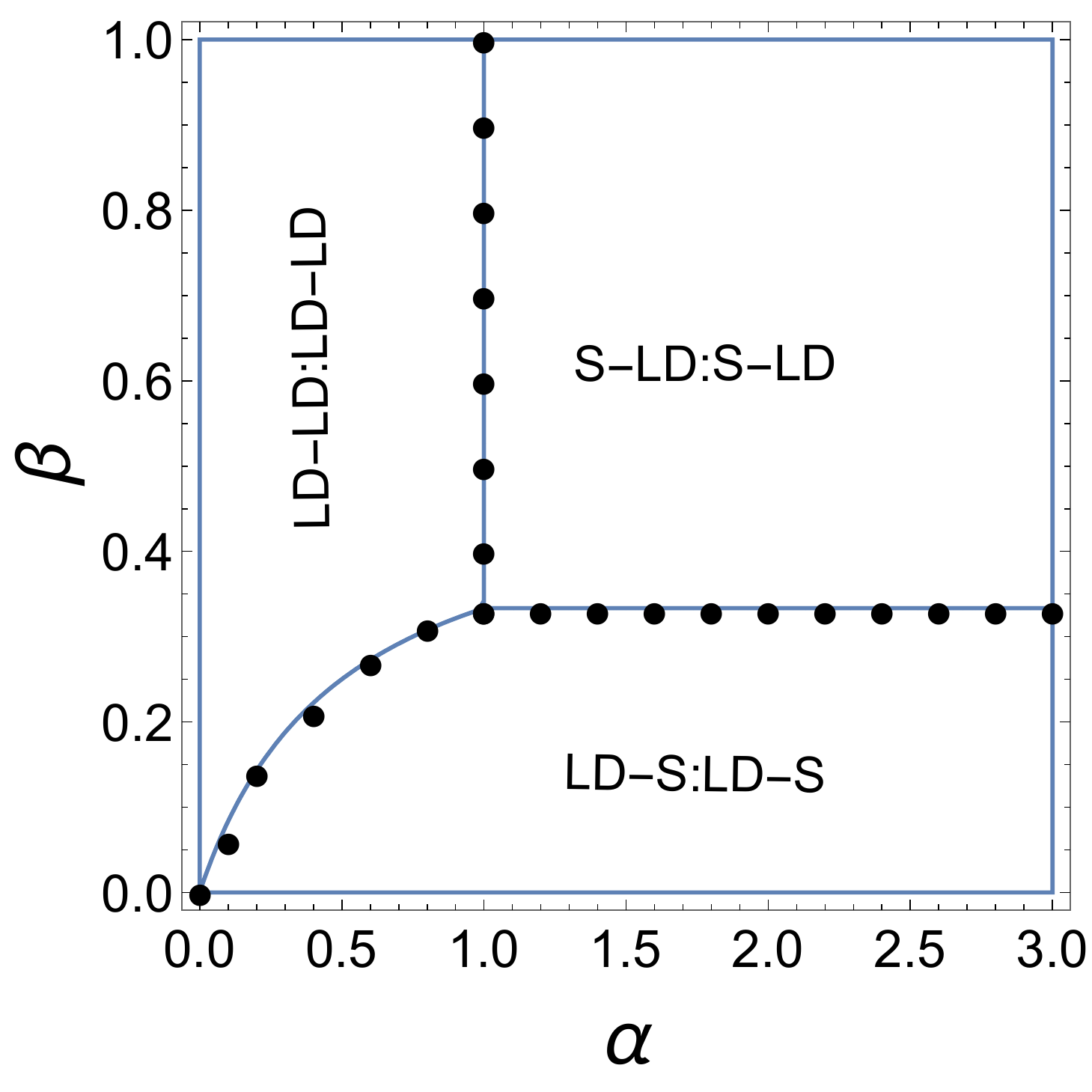}}\\
\subfigure[\label{12}$\mu=1.2$]{\includegraphics[width = 0.3\textwidth]{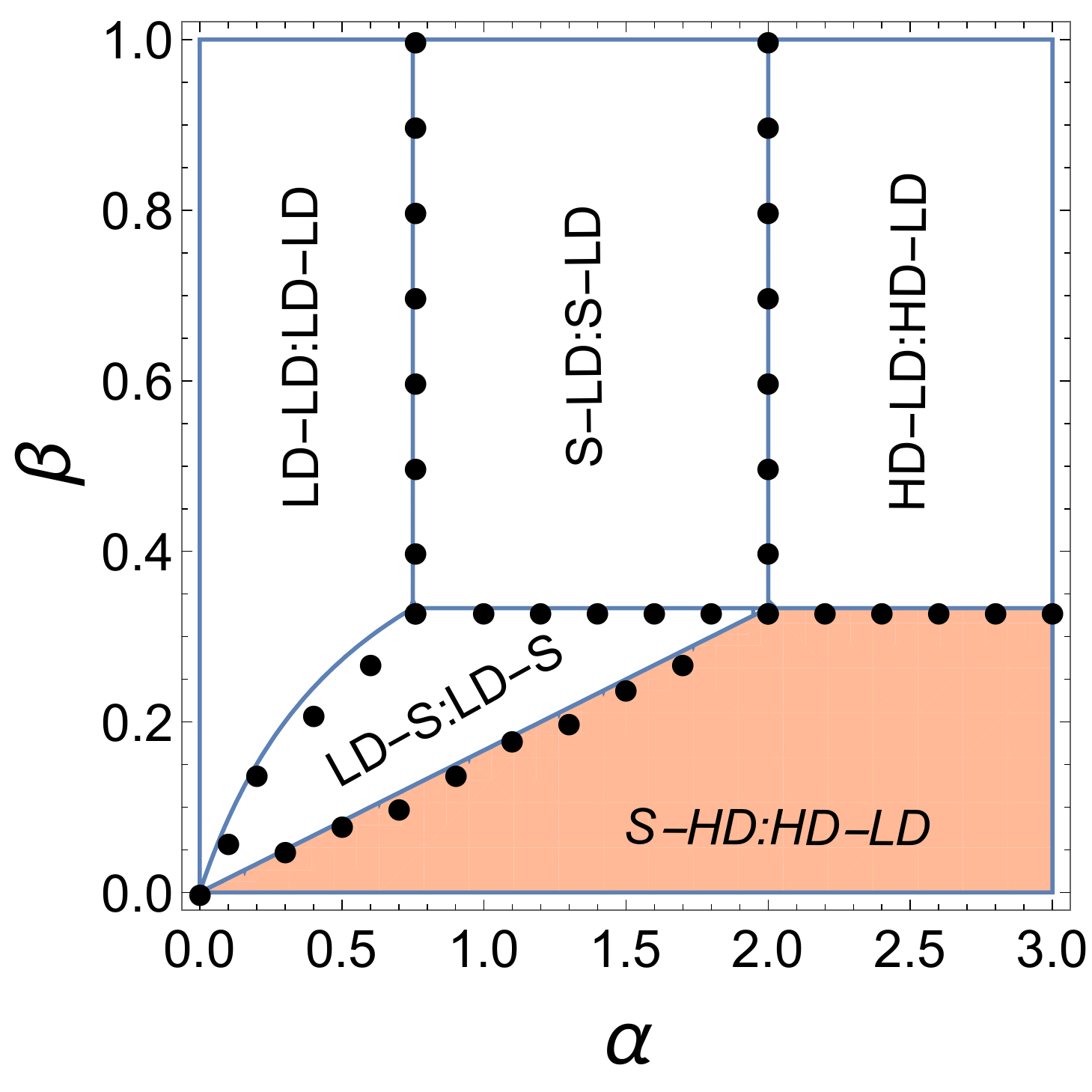}}
\subfigure[\label{15}$\mu=1.5$]{\includegraphics[width = 0.3\textwidth]{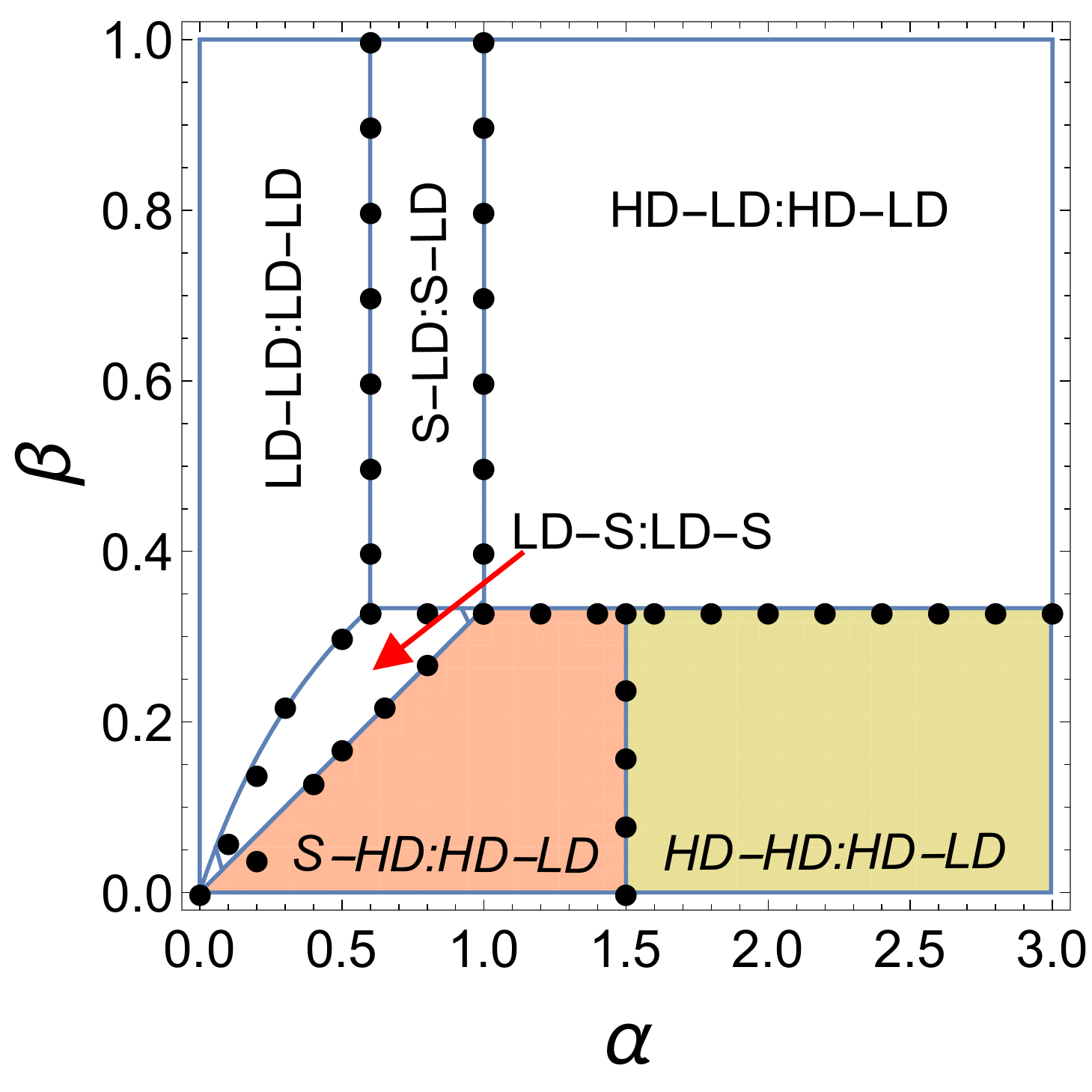}}
\subfigure[\label{2}$\mu=2$]{\includegraphics[width = 0.3\textwidth]{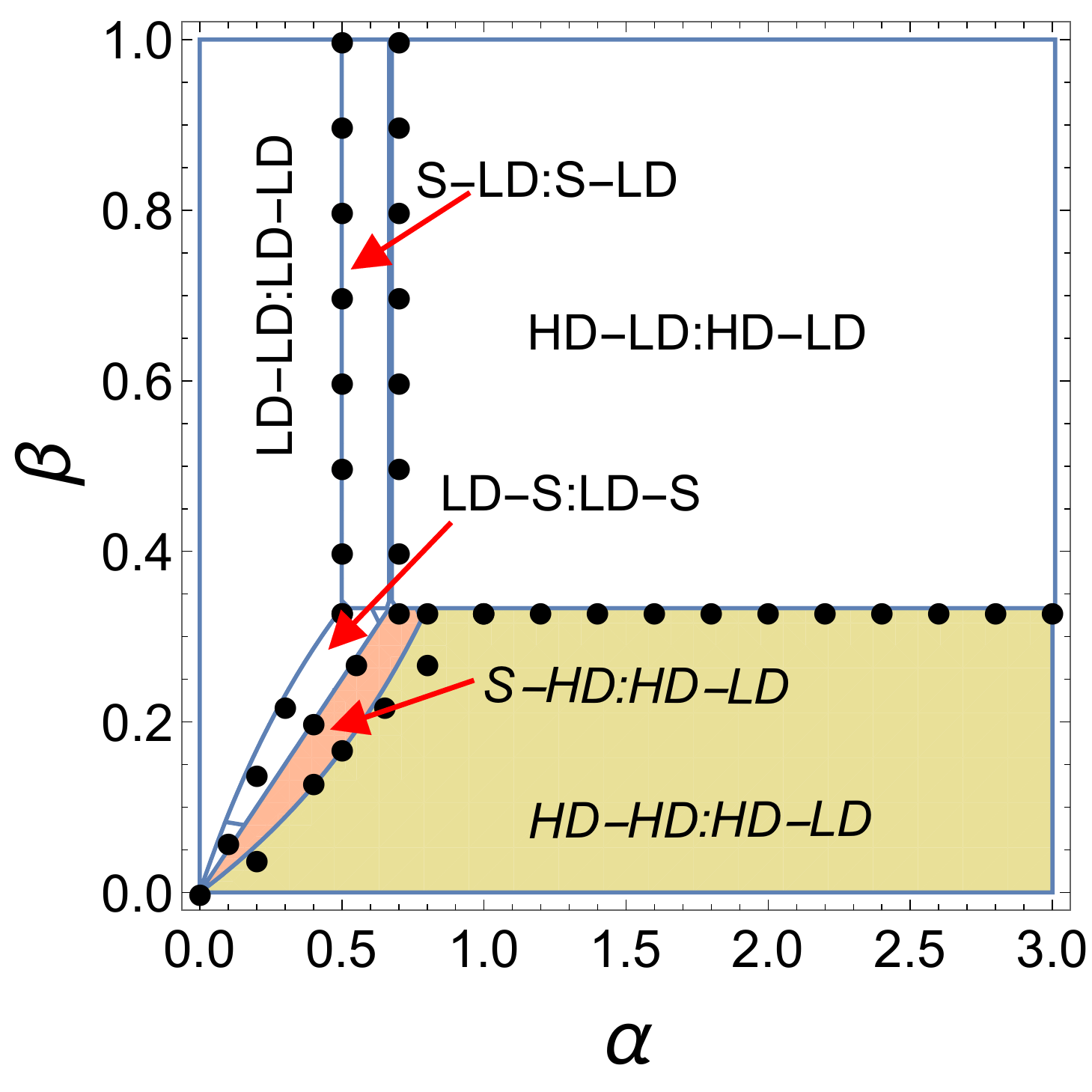}}
\caption{\label{Sphases} Stationary phase diagrams for increasing parameter $\mu=N_{tot}/N$ notified in sub-captions of (a-f). In the limiting case $\mu\rightarrow\infty$, the phase diagram converges to the two intersected lane model with infinite particles \cite{tian2020totally}. The white coloured regions represent symmetric phases, whereas, the coloured regions denote asymmetric phases. The black circles are the simulated results. The phase boundaries are computed
within an estimated error of less than $1\%$.}
\end{figure*}
\section{Stationary system states}\label{phases}
In this section we elaborate the qualitative and quantitative behaviour of stationary phase diagrams depending on three controlling parameters $(\alpha,\beta,\mu)$. As already discussed, we have divided each lane into two segments the possible phases are labelled as A-B:C-D where A and B describes the phase in left and right segment of lane $L$. Similarly, C and D symbolises the phase manifested in $T_1$ and $T_2$. In addition, a phase is characterised as a symmetric phase if the particle density in $L_1(L_2)$ is equal to density in $T_1(T_2)$ (i.e. $\rho_1^b=\sigma_1^b$ and $\rho_2^b=\sigma_2^b$). Otherwise, the phase is indicated as a asymmetric phase and is labelled in \textit{italics}.

For the proposed model, each segment can exhibit four possible stationary states LD, HD, MC or S. Therefore, the maximum possible number of stationary phases in each lane is $2^4=16$. However, the existence of an ample number of phases is prohibited due to various restrictions. For example, from Eqn. \eqref{current} it can be easily realised that for either lane the possibility of having MC phase in any of the segments and the LD, HD or S in the other segment can be discarded because these phases support different particle currents. Moreover, both the segments cannot exhibit MC phase simultaneously. This is because if left segment shows average density 1/2, the inhomogeneous dynamical rules does not allow the right segment to achieve maximal current. 
Also, due to interaction of particles at site $k$, the right segments of both the lanes cannot exhibit HD phase simultaneously. Since, a particle at site $k$ is restricted to jump only in the same lane. As a consequence, when any of particle arriving from lane $L$ or $T$ resides on the intersected site, the particle on site $k-1$ of other lane has to wait till the particle at $k^{th}$ site jumps to the site of its own respective lane.  Owing to this fact, a asymmetric phase has been observed in ref.\cite{tian2020totally}.

Now we theoretically investigate the conditions of existence of phases, phase boundaries for the proposed model with varying entry and exit rates. We provide explicit expressions for the density profiles, phase boundaries and position of shock in terms of $\mu$.

\subsection{Symmetric Phases}
We now discuss the occurrence of different symmetric phases and aim to calculate the effective rates and densities to determine the phase boundaries. As discussed, for symmetric phase the following conditions hold i.e.,
\begin{eqnarray}\label{symcon1}
\rho_1^b=\sigma_1^b,\qquad \rho_2^b=\sigma_2^b,
\end{eqnarray} 
that also implies,
\begin{eqnarray}
\label{symcon2} 
\alpha_{eff,L}=\alpha_{eff,T}=\alpha_{eff}, \quad \rho^{k}=\sigma^{k}.
\end{eqnarray} 
Without any loss of generality, we can thus analyse the dynamics of particles in any one lane (say $L$). The same results are pertinent for the other lane $T$. Hence, Eqn. \eqref{finite} reduces to,
\begin{equation}\label{finitesym}
\alpha^* =\alpha-\dfrac{2\alpha}{\mu}\left(\int_0^{1/2}\rho_1^b~dx+\int_{1/2}^{1}\rho_2^b~dx\right).
\end{equation}
To determine the conditions of existence of different phases we need to compute effective rates $\alpha^*$, $\alpha_{eff}$ and $\beta_{eff}$. Depending on the phase we can also determine the densities, $\rho_1^b$, $\rho_2^b$, $\rho^{k}$, $\rho_1^{k-1}$ and $\rho_2^{k+1}$.
\subsubsection{LD-LD:LD-LD Phase}
In this phase, we assume both the segments of lane $L$ to exhibit low-density phase. Each homogeneous segment is entry-dominated for which the conditions of existence are given by,

\begin{eqnarray}
\alpha^*<\min\{\beta_{eff},1/2\},\qquad
\alpha_{eff}<\min\{\beta,1/2\}.
\end{eqnarray}
The density of particles in the bulk of each segment is given by,
\begin{eqnarray}\label{int}
\rho^1=\rho_1^b=\alpha^* \quad and \quad \rho_2^{k+1}=\rho_2^b=\alpha_{eff}.
\end{eqnarray}
Since, current is equal for both the segments Eqn. \eqref{current} results in,
\begin{eqnarray}
\alpha^* &=&\alpha_{eff}.
\end{eqnarray}
Solving Eqn. \eqref{finitesym}, we have,
\begin{eqnarray}
\alpha^* =\alpha-\dfrac{2\alpha}{\mu}\int_0^{1}\alpha^*~dx,
\end{eqnarray}
that leads to
\begin{eqnarray}\label{alphald}
\alpha^* =\dfrac{\alpha \mu}{\mu+2\alpha}.
\end{eqnarray}
The density of particles arriving from lane $L$ at intersected site is given by Eqn. \eqref{inter}, $\rho^{k}=\alpha^*$.
The effective exit rate with which a particle leaves from left segment given in Eqn. \eqref{ebeta} reduces to,
\begin{equation}
\beta_{eff}=1-2\alpha^*.
\end{equation}
In addition, stationary current argument at site $k-1$ implies that the bulk current in right segment is equal to the current entering into it that yields,
\begin{eqnarray}
\rho_1^{k-1}=\dfrac{\alpha^*(1-\alpha^*)}{1-2\alpha^*}.
\end{eqnarray} 
The conditions of existence for this phase thus reduces to,
\begin{eqnarray}
\alpha^*<\min\{\beta,1/3\}.
\end{eqnarray}
For the case when $\mu\rightarrow\infty$, the expression for effective intrinsic rate in Eqn. \eqref{alphald} reduces to $\alpha^*=\alpha$. As a result, we retrieve the phase boundaries for the model with infinite number of particles given by $\alpha<\min\{\beta,1/3\}$.
\subsubsection{HD-LD:HD-LD Phase}
We assume in this phase for each lane, left segment to exhibit high density phase and right segment in low density phase. Correspondingly, the homogeneous left segment is exit-dominated, whereas, right segment is entry dominated. This phase is determined by,
\begin{eqnarray}
\beta_{eff}<\min\{\alpha^*,1/2\},\qquad
\alpha_{eff}<\min\{\beta,1/2\}.
\end{eqnarray}
The density of particles in the bulk of each segment is given by,
\begin{eqnarray}
\rho_1^b=\rho_1^{k-1}=1-\beta_{eff} \quad and \quad \rho_2^b=\rho_2^{k+1}=\alpha_{eff}.
\end{eqnarray}
The condition of constant current in Eq. \eqref{current} leads to,
\begin{eqnarray}
\beta_{eff} &=&\alpha_{eff}.
\end{eqnarray}
Clearly, Eqn. \eqref{inter} provides the density of particles at site $k$ as $\rho^{k}=\beta_{eff}$. Now, plugging these values in Eqn. \eqref{ebeta} we obtain,
\begin{eqnarray}
\beta_{eff}=1-2\beta_{eff},
\end{eqnarray}
implying,
\begin{eqnarray}
\beta_{eff}=\dfrac{1}{3}.
\end{eqnarray}
Further, the effective intrinsic rate given in Eqn. \eqref{finitesym} is obtained as follows,
\begin{eqnarray}
\alpha^* = \alpha-\dfrac{2\alpha}{\mu}\left(\int_0^{1/2}(1-\beta_{eff})~dx+\int_{1/2}^1\alpha_{eff}\right),
\end{eqnarray}
that yields,
\begin{eqnarray}\label{muhlhl}
\alpha^* =\alpha\left(\dfrac{\mu-1}{\mu}\right).
\end{eqnarray}
From the above equation we conclude that this phase exists only when $\mu>1$. The conditions of existence reduces to,
\begin{equation}\label{conhl}
1/3<\min\{\alpha^*,\beta\}.
\end{equation}
In the limiting case, as $\mu\rightarrow \infty$, the effective intrinsic rate in Eqn. \eqref{muhlhl} reduces to $\alpha$. Correspondingly, the conditions that favor the existence of this phase reduces to $1/3<\min\{\alpha,\beta\}$ for the model with infinite number of particles  \cite{tian2020totally}.
\subsubsection{S-LD:S-LD Phase}
For this phase, the particles in $L_1$ exhibit shock phase i.e. a part of segment is in LD phase and the rest is in HD phase. Whereas, the right segment shows low-density phase. The existence of this phase is determined by following conditions,
\begin{eqnarray}
\alpha^*=\beta_{eff},~~\beta_{eff}<1/2,\qquad
\alpha_{eff}<\min\{\beta,1/2\}.
\end{eqnarray}
For this phase, in left segment the site $1$ and $k-1$ is entry and exit-dominated respectively. The density in $L_1$ is written as,
\begin{eqnarray}
\rho^1&=&\alpha^*,\\
\rho_1^{k-1} &=& 1-\beta_{eff},\\
\int_0^{1/2}\rho_1^b~dx &=&\int_0^{x_w}\alpha^*~dx+\int_{x_w}^{1/2}(1-\beta_{eff})~dx.
\end{eqnarray}
Similarly, the density in $L_2$ is given by,
\begin{equation}
\rho_2^b=\rho_2^{k+1}=\alpha_{eff}.
\end{equation}
Also, the current is constant in both the segments as given in Eqn. \eqref{current} that leads to,
\begin{equation}\label{rates}
\alpha^*=\beta_{eff}=\alpha_{eff}.
\end{equation}
From Eqn. \eqref{ebeta} we obtain the effective exit rate of particles from left segment as,
\begin{equation} 
\beta_{eff}=\frac{1}{3}.
\end{equation}
Hence, Eqn. \eqref{finitesym} yields,
\begin{equation}
\begin{split}
\alpha^* =\alpha-\dfrac{2\alpha}{\mu}\left(\int_0^{x_w}\alpha^*~dx+\int_{x_w}^{1/2}(1-\beta_{eff})~dx\right.\\+\left.\int_{1/2}^{1}\alpha^*~dx\right),
\end{split}
\end{equation}
that reduces to,
\begin{eqnarray}
\alpha^* =\alpha\left(\dfrac{3(\mu-1)+2x_w-1}{3\mu}\right).
\end{eqnarray}
Since, $\alpha^*=\frac{1}{3}$ from Eqn. \eqref{rates}, the shock position is given by,
\begin{equation}\label{ssld}
x_w=\dfrac{3}{2}\left(\dfrac{\mu}{3\alpha}-\mu+1\right).
\end{equation}
As the shock position is bounded between $0<x_w<1/2$ that provides one of the condition for existence of this phase. This $x_w$ depends on the parameter $\mu$ and $\alpha$. For a fixed value of $\mu$, as $\alpha$ increases shock travels to left of the lattice on left segments of lane $L$. Hence, this phase exists when,
\begin{equation}
0<x_w<1/2,\qquad \beta\geq 1/3.
\end{equation}
For a system with infinite number of particles $\mu\rightarrow\infty$, we can clearly see that $x_w\rightarrow\infty$, as a result this phase cease to exist and converges to HD-LD:HD-LD phase \cite{tian2020totally}. 
\subsubsection{LD-S:LD-S Phase}
In each lane, the density of particles in left segment is in low density phase, while, in right segment particles are in shock phase. The conditions that support the existence of this phase are,
\begin{eqnarray}
\alpha^{*}<\min\{\beta_{eff},1/2\},\qquad
\alpha_{eff}=\beta,~~\beta<1/2.
\end{eqnarray}
Since, the right segment is entry-dominated, we can write the density at site $k+1$, $\rho_2^{k+1}=\alpha_{eff}$. 
By the current constancy condition from Eqn. \eqref{current}, the rates are given by,
\begin{equation}
\alpha^*=\alpha_{eff}=\beta.
\end{equation}
The density of particles in left and right segment is given by, 
\begin{eqnarray}
\rho_1^b &=& \rho^1=\alpha^*,\qquad \rho_2^{k+1}=\alpha_{eff},\\
\int_{1/2}^1 \rho_2^b ~dx &=&\int_{1/2}^{x_w} \alpha_{eff} ~dx + \int_{x_w}^1 (1-\beta) ~dx,
\end{eqnarray}
respectively. 
As a result, solving Eq. \eqref{finitesym} provides the effective intrinsic rate given by,
\begin{eqnarray}
\beta &=& \alpha-\dfrac{2\alpha}{\mu}\left(\int_0^{1/2} \beta ~dx-\int_{1/2}^{x_w} \beta ~dx-\int_{x_w}^{1} (1-\beta) ~dx\right),\qquad
\end{eqnarray}
that implies
\begin{eqnarray}\label{slds}
x_w &=& \dfrac{\beta\mu-\alpha(\mu+2\beta-2)}{2\alpha(1-2\beta)}.
\end{eqnarray}
For this phase to exist the shock travels within the range $1/2<x_w<1$. Also, from Eqn. \eqref{ebeta} the density at $k^{th}$ site is $\rho^k=\alpha_{eff}=\beta$. Hence, the effective exit rate is given as,
\begin{eqnarray}
\beta_{eff} &=& 1-2\beta.
\end{eqnarray}
In addition, stationary current argument at site $k-1$ implies that the bulk current in right segment is equal to the current entering into it that yields,
\begin{eqnarray}
\rho_1^{k-1}=\dfrac{\alpha^*(1-\alpha^*)}{1-2\beta}.
\end{eqnarray}
The conditions of this phase to exist are,
\begin{equation}
1/2<x_w<1, \qquad \beta\leq 1/3.
\end{equation}
For $\mu\rightarrow \infty$, the computed expression for the position of shock converges to $\infty$. This shows that this symmetric phase vanishes when there are infinite number of particles and tends to LD-LD:LD-LD phase \cite{tian2020totally}.
\begin{figure*}[!htb]
\centering
\subfigure[\label{lsls}$\alpha=0.5,~\beta=0.2$]{\includegraphics[width = 0.45\textwidth]{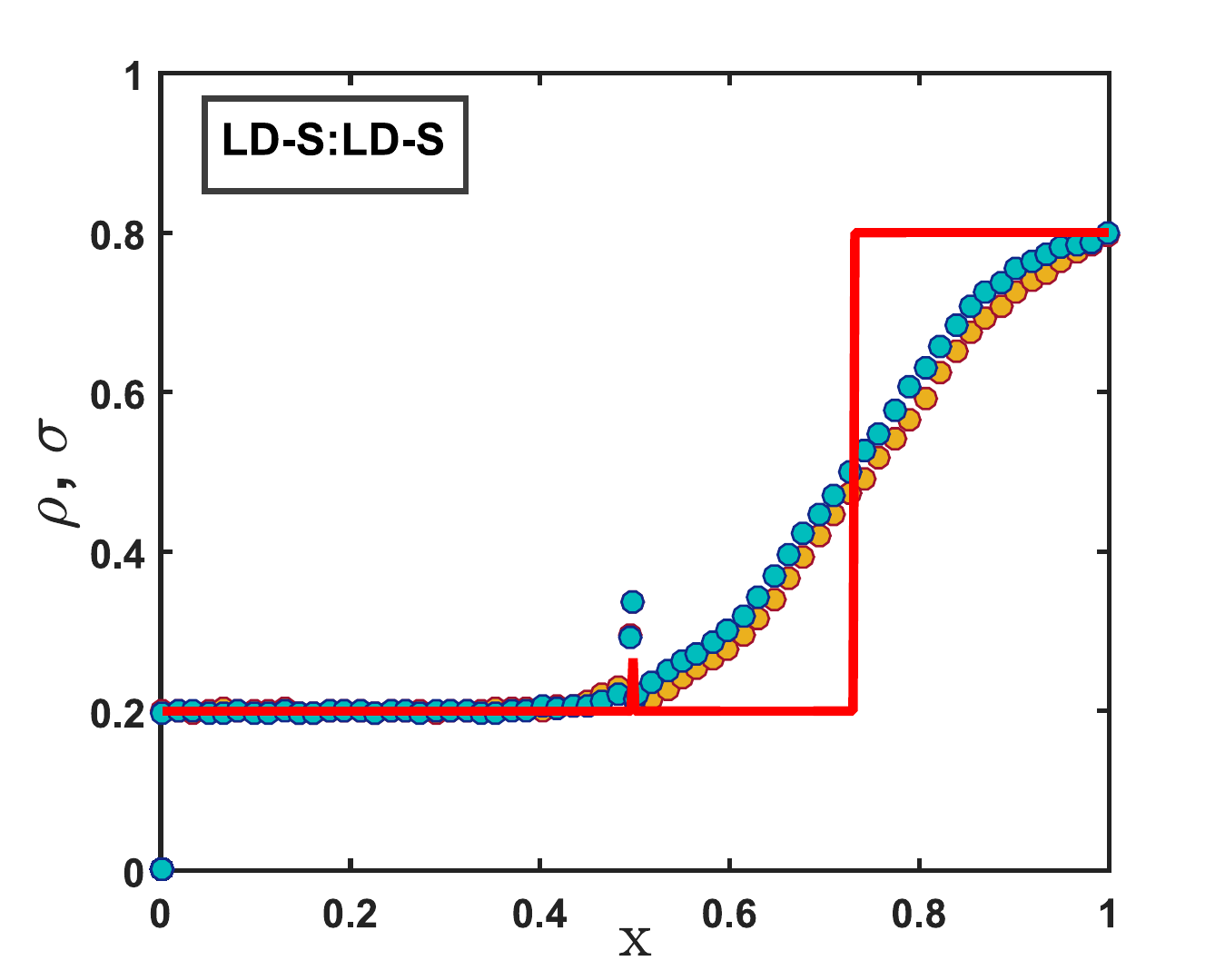}}
\subfigure[\label{ldld}$\alpha=0.1,~\beta=0.8$]{\includegraphics[width = 0.45\textwidth]{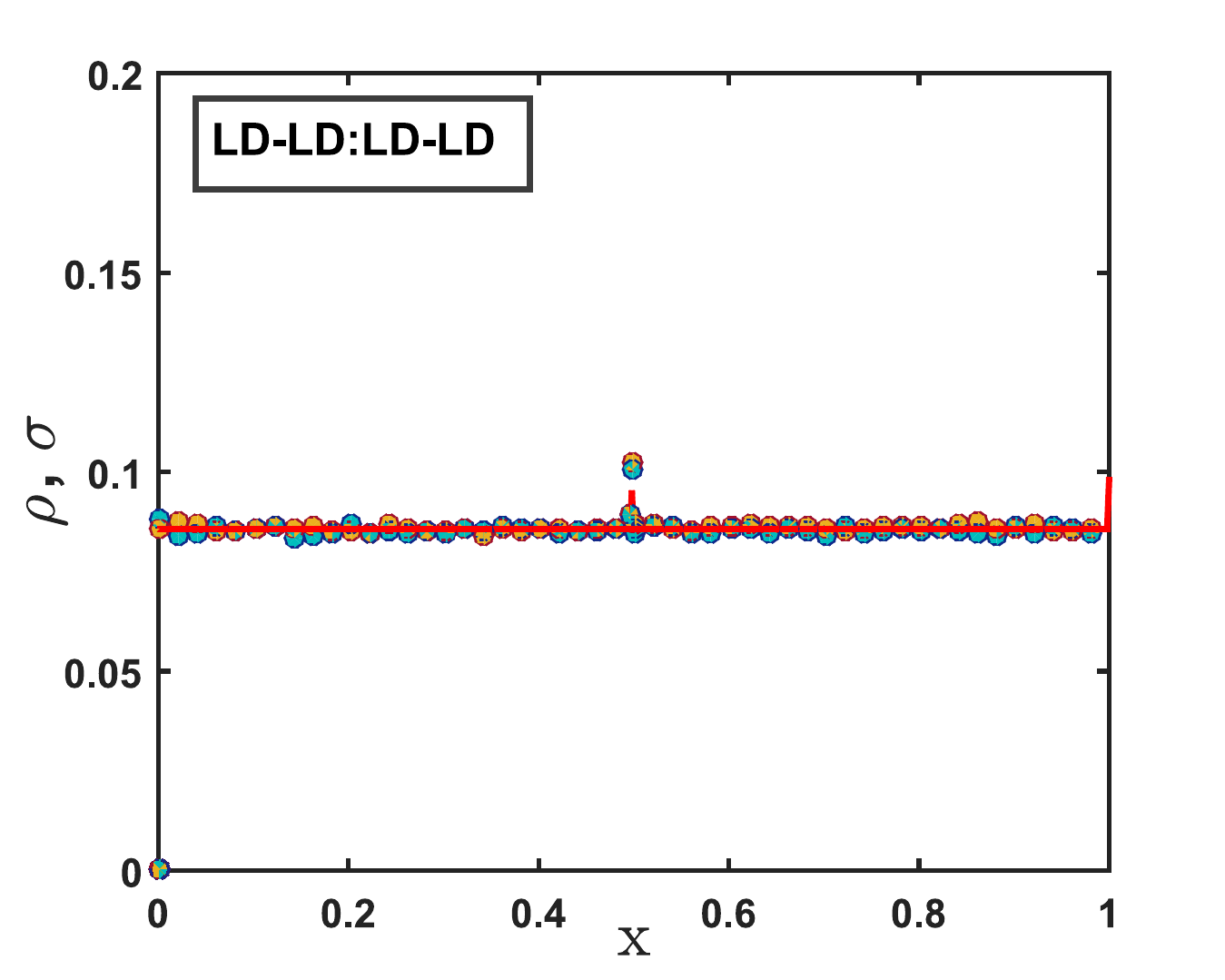}}\\
\subfigure[\label{hdldhdld}$\alpha=2.5,~\beta=0.8$]{\includegraphics[width = 0.45\textwidth]{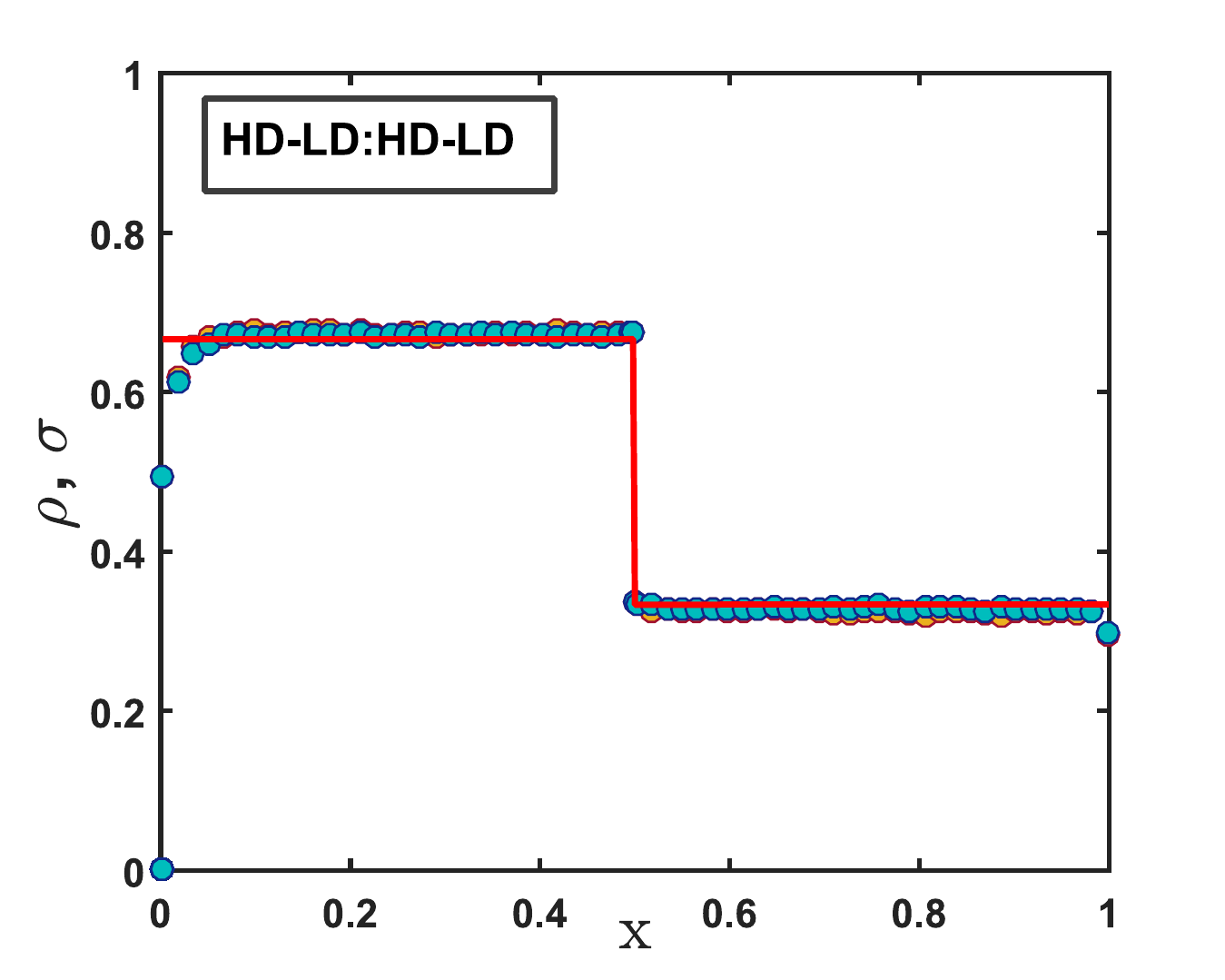}}
\subfigure[\label{slsl}$\alpha=1.5,~\beta=0.8$]{\includegraphics[width = 0.45\textwidth]{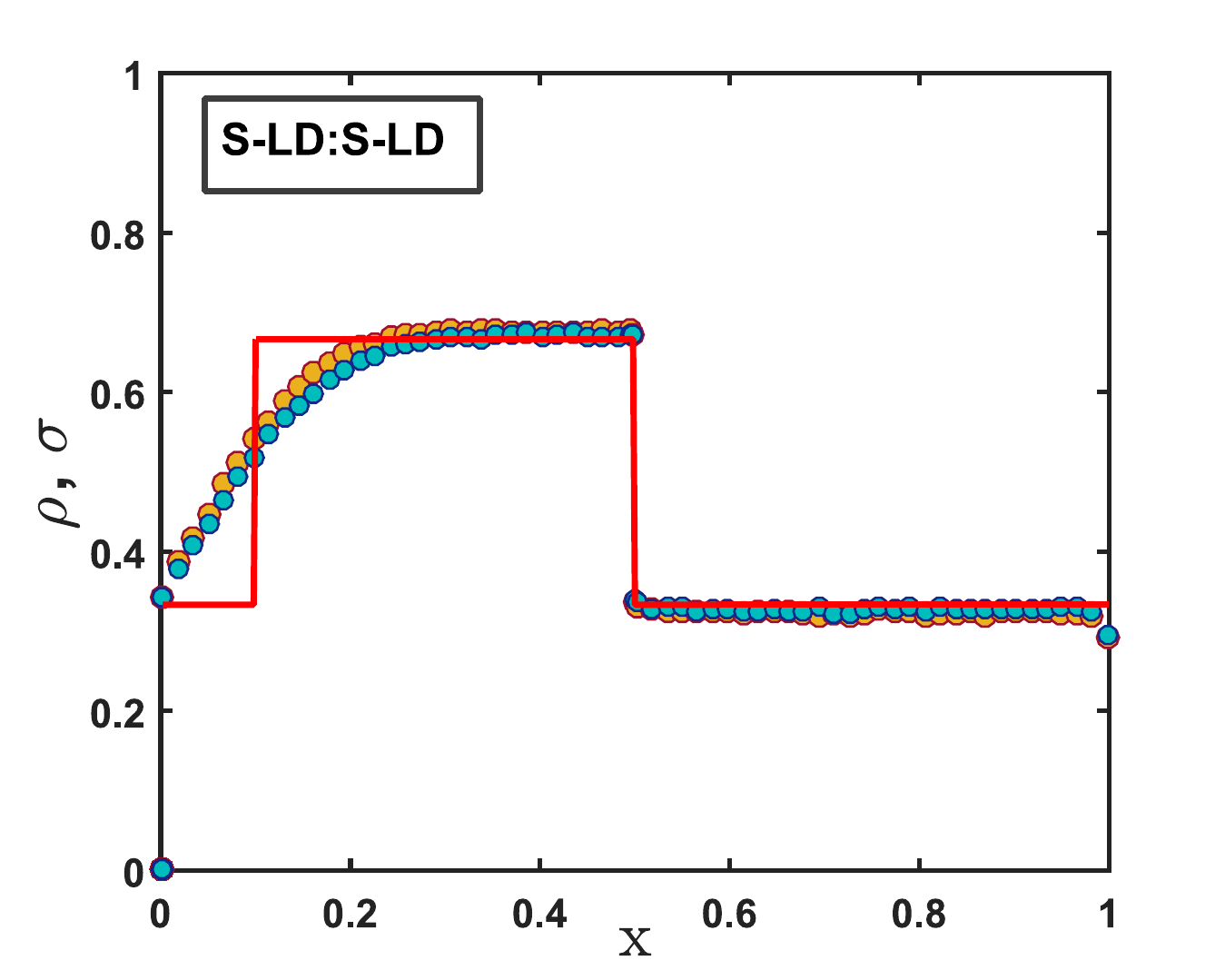}}
\caption{\label{densym}Density profiles attributed to symmetric phases for $(\alpha,\beta)$ notified in sub-captions of (a-d) with parameter $\mu=1.2$. Solid lines represent theoretical results, while markers denote the simulated results.}
\end{figure*}
\begin{figure*}[!htb]
\centering
\subfigure[\label{hdhdhdld}]{\includegraphics[width = 0.35\textwidth]{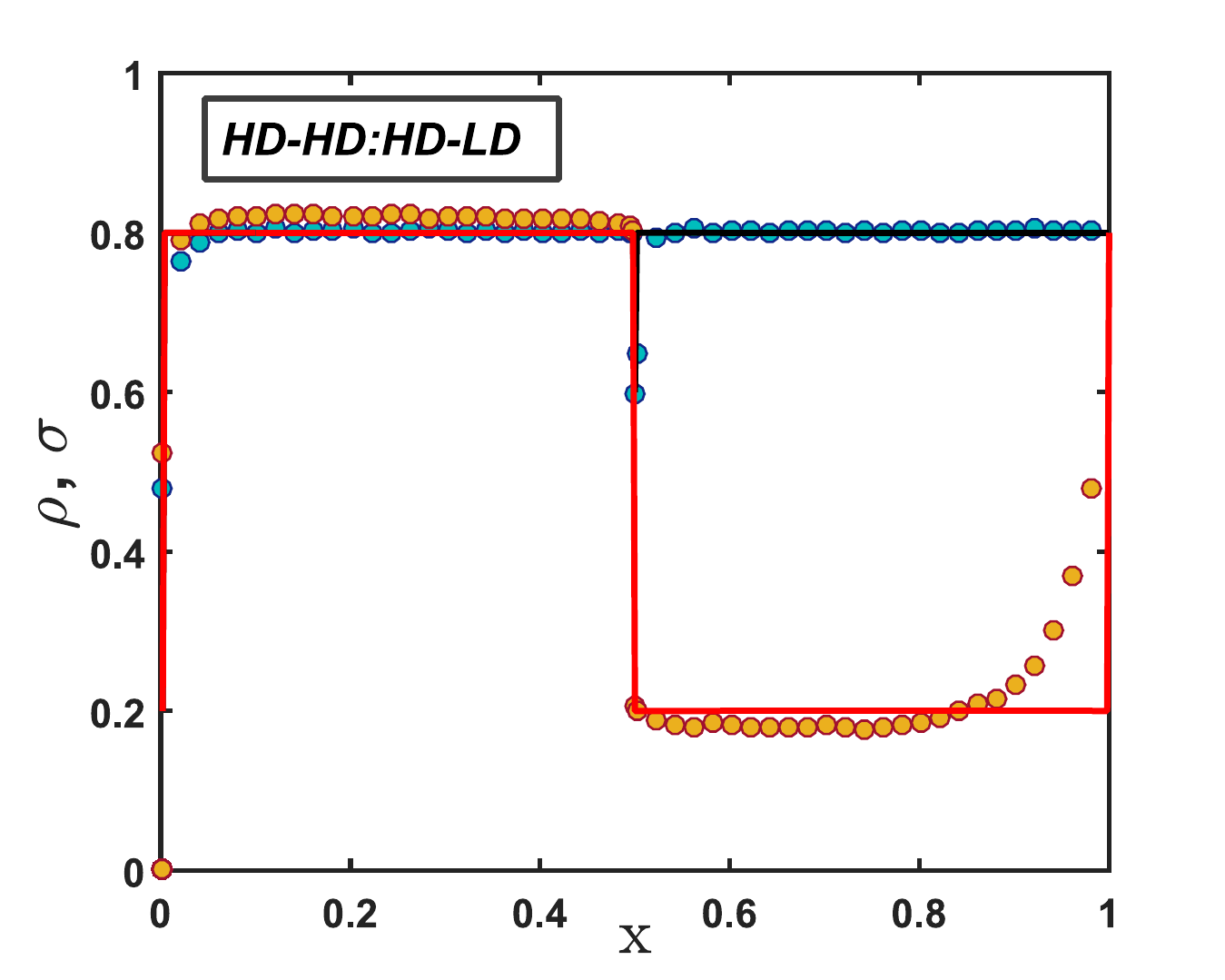}}
\subfigure[\label{hishl}]{\includegraphics[width = 0.4\textwidth]{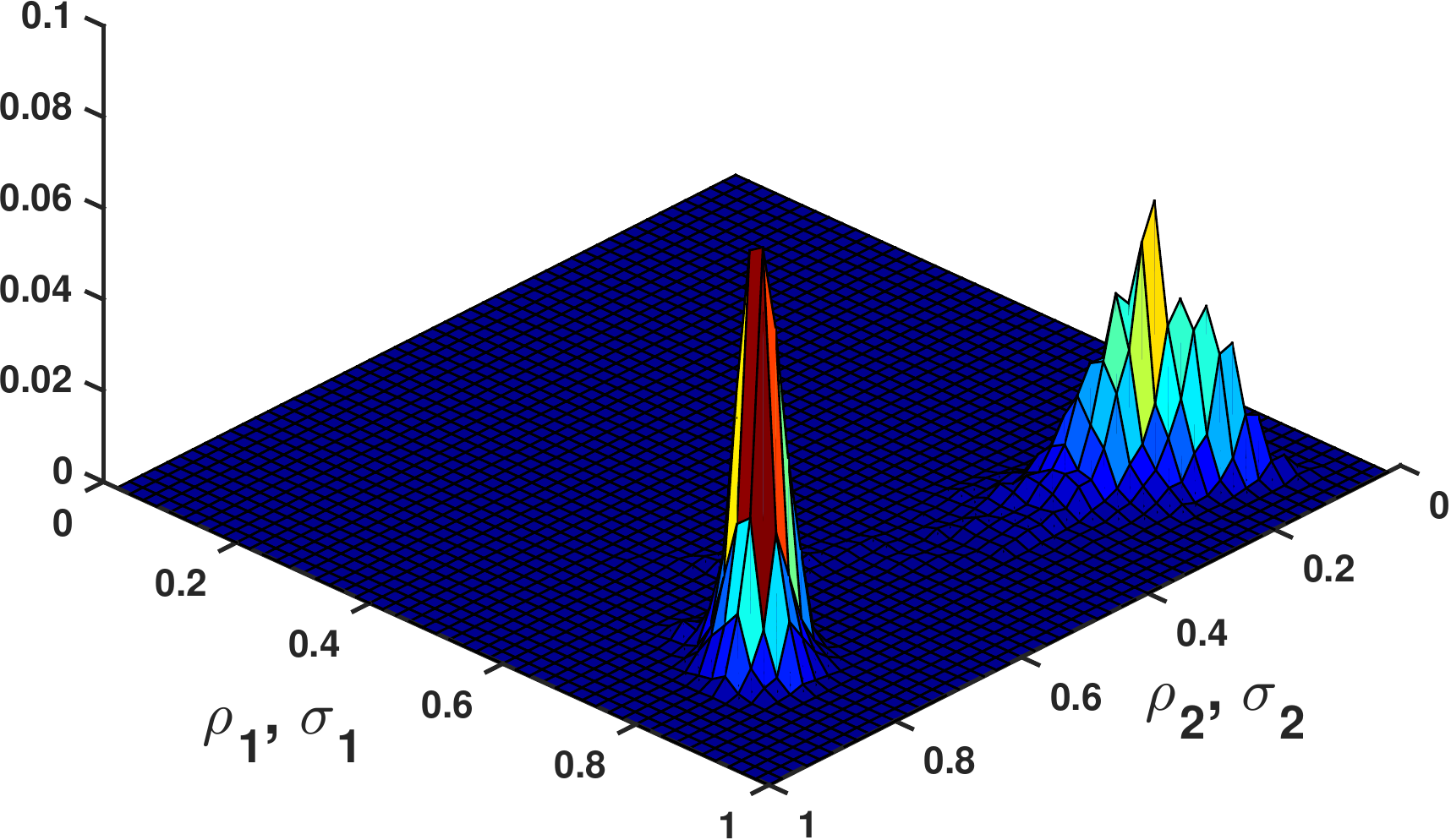}}\\
\subfigure[\label{shd}]{\includegraphics[width = 0.35\textwidth]{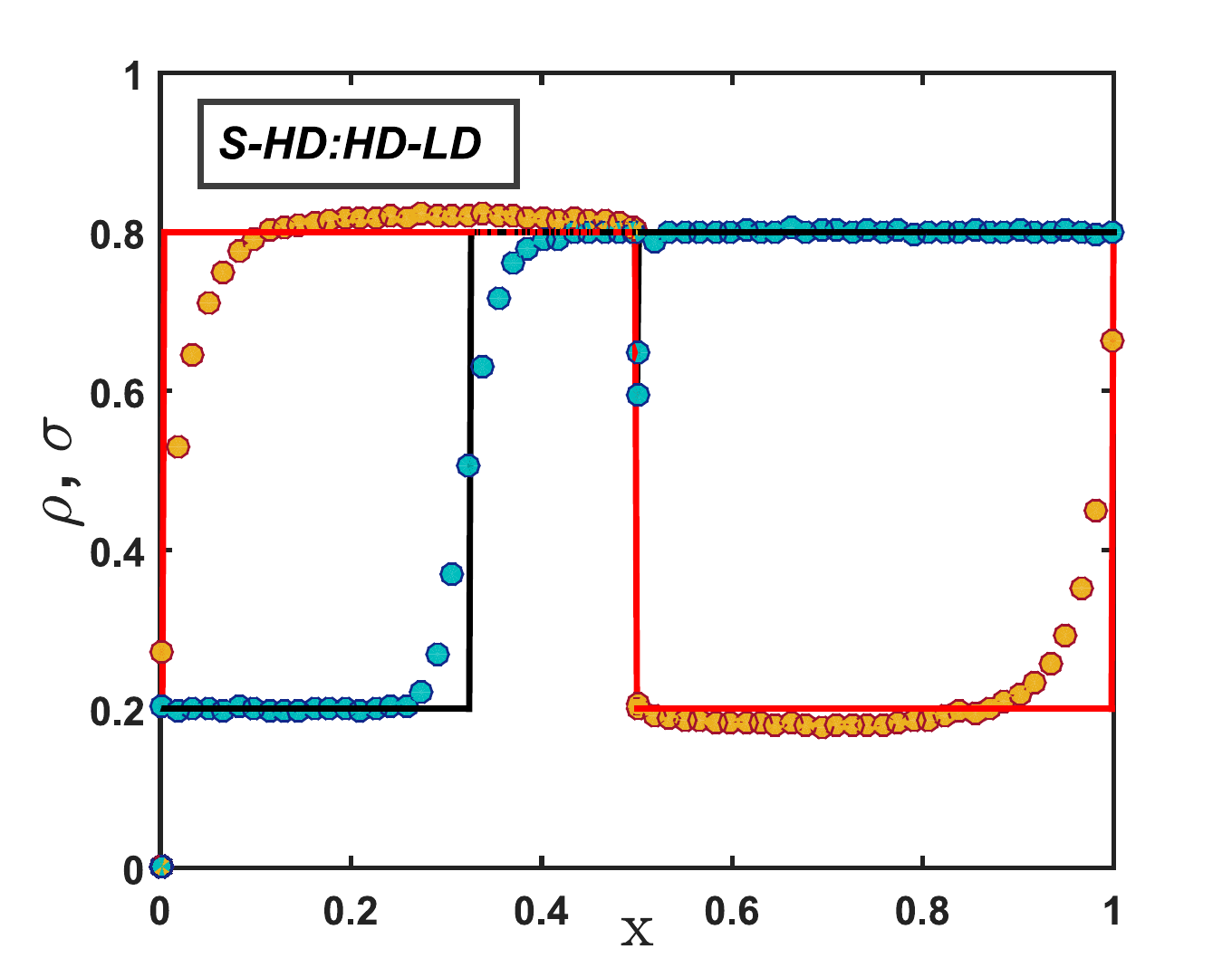}}
\subfigure[\label{hisshd}]{\includegraphics[width = 0.4\textwidth]{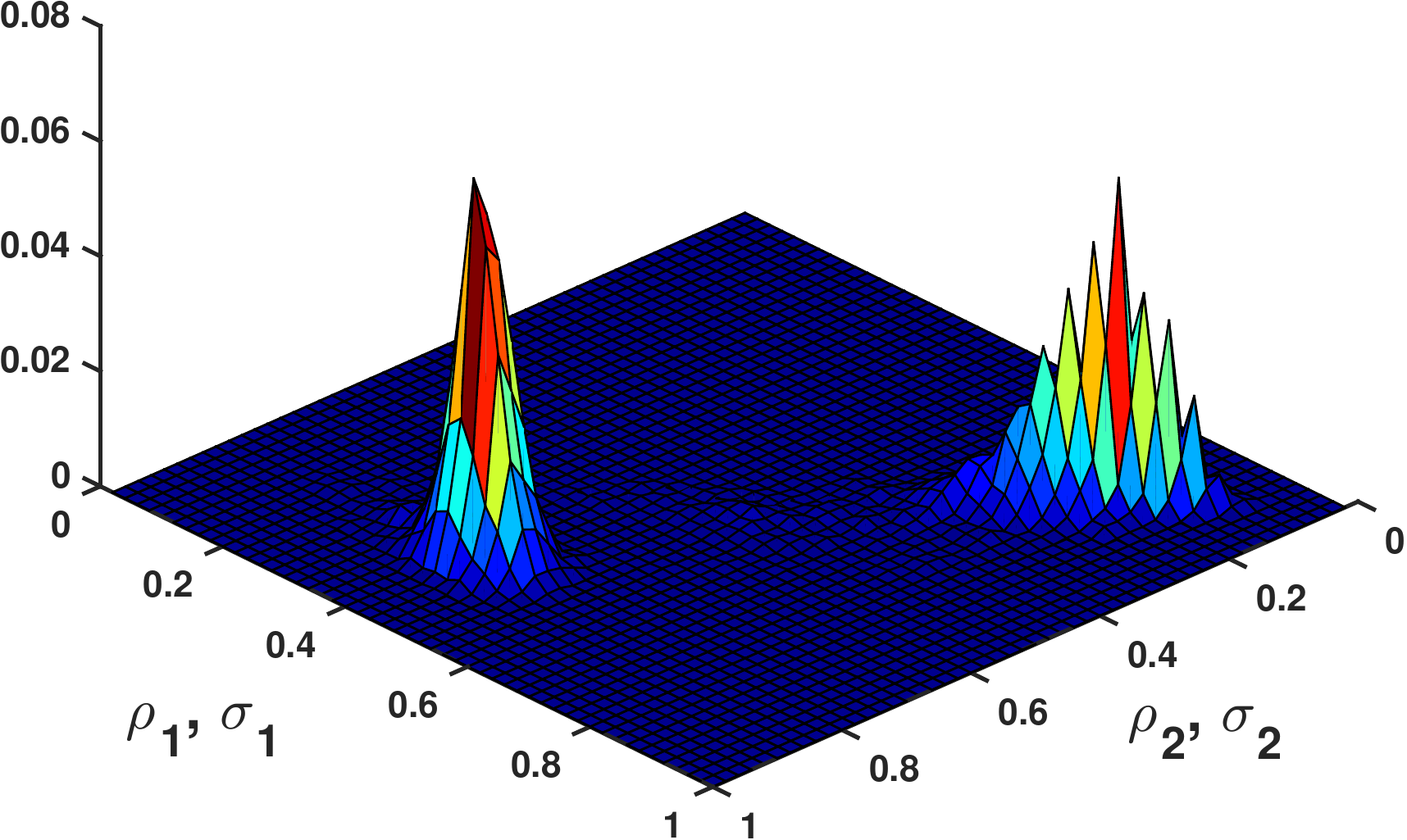}}
\caption{\label{denasym}(Left) Density profiles attributed to asymmetric phases for $\alpha=2.5,~\beta=0.2$ in (a) and (c) with parameter $\mu=1.5$ and $\mu=1.2$ respectively. Such parameters are chosen to show how for a particular value of $\alpha,~\beta$, considering different values of $\mu$, two different asymmetric phases exist. Solid lines represent theoretical results, while markers denote the simulated results. (Right) Particle-density histograms (b) and (d) corresponding to (a) and (c) respectively are plotted to probe the phenomenon of symmetry breaking through Monte Carlo simulations.}
\end{figure*}

\subsection{Asymmetric Phases}
For asymmetric phase, the characteristics of both the lanes are different. So, we need to consider densities of both the lanes $L$ and $T$ in our further calculations. 
The specifications that support the extant of asymmetric phases are,
\begin{eqnarray}
\rho_1^b\neq\sigma_1^b,\qquad\text{or}\qquad \rho_2^b\neq\sigma_2^b,
\end{eqnarray}
that can lead to
\begin{eqnarray}
\alpha_{eff,L}\neq\alpha_{eff,T}, \quad \rho^{k}\neq\sigma^{k}.
\end{eqnarray}
We aim to explicitly calculate the effective rates for both the lanes and densities in bulk as well as at intersected site of particles moving in each lane. 

\subsubsection{\textit{S-HD:HD-LD} Phase}
We presume particles in left segment of lane $L$ to exhibit shock phase and right segment to manifest high density phase. While, the particles in left and right segment of lane $T$ display high and low density phase respectively. This phase exists when the boundary controlling parameters satisfies the following conditions described in table \eqref{sh}.
\begin{table}[!htb]
\centering
\caption{\label{sh} Conditions for existence of asymmetric \textit{S-HD:HD-LD} Phase}
\begin{tabular}{|c||c||c|}
\hline
& $L$ & $T$\\
\hline
\hline
Left segment & $\alpha^*=\beta_{eff},~~ \beta_{eff}<1/2$ & $\beta_{eff}<\min\{\alpha^*,1/2\}$\\
Right segment & $\beta<\min\{\alpha_{eff},1/2\}$ & $\alpha_{eff,T}<\min\{\beta,1/2\}$.\\
\hline
\end{tabular}
\end{table}\\
The density in bulk of two segments of lane $L$ is given by,
\begin{eqnarray}
\rho_1^{k-1} &=& 1-\beta_{eff},\\
\int_0^{1/2}\rho_1^b~dx &=&\int_0^{x_w}\alpha^*~dx+\int_{x_w}^{1/2}(1-\beta_{eff})~dx,\\
\rho_2^b &=& 1-\beta.
\end{eqnarray}
Next, the density in the bulk of left and right segment of lane $T$ is given by,
\begin{eqnarray}
\label{asymh1}
\sigma_1^{k-1}=\sigma_1^b &=& 1-\beta_{eff},\\\label{asymh2}
\sigma_2^{k+1}=\sigma_2^b &=& \alpha_{eff,T},
\end{eqnarray}
respectively. Since, current is constant in each segment of $L$, from Eqn. \eqref{eqcur} we have,
\begin{equation}
\alpha^*(1-\alpha^*)=\beta_{eff}(1-\beta_{eff})=\beta(1-\beta)
\end{equation}
and similarly for lane $T$, we can write,
\begin{equation}
\beta_{eff}(1-\beta_{eff})=\alpha_{eff,T}(1-\alpha_{eff,T}).
\end{equation}
This yields,
\begin{equation}
\alpha^*=\alpha_{eff,T}=\beta
\end{equation}
and these values are further substituted in Eqn. \eqref{finite} to evaluate the shock position in $L_1$. 
\begin{equation}
\begin{split}
\beta =\alpha-\dfrac{\alpha}{\mu}\left(\int_0^{x_w}\beta ~dx+\int_{x_w}^{1/2}(1-\beta)~dx\right.\\ +\left.\int_0^{1}(1-\beta)~dx+\int_{1/2}^1 \beta ~dx\right).
\end{split}
\end{equation}
The shock position $x_w$ is obtained as,
\begin{equation}\label{sshd}
x_w=\dfrac{2\mu(\beta-\alpha)+\alpha(3-2\beta)}{2\alpha(1-2\beta)}
\end{equation}
that lies in [0,1/2]. From Eqn. \eqref{ebeta}, plugging $\sigma^{k}=\alpha_{eff,T}=\beta$ we have,
\begin{eqnarray}
\beta_{eff}=1-\rho^{k}-\beta,
\end{eqnarray}
that provides the density of particles arriving from lane $L$ at intersected site, given by,
\begin{eqnarray}
\rho^{k}= 1-2\beta.
\end{eqnarray}
This gives the effective entry rate of particles in $L_2$, $\alpha_{eff,L}=1-2\beta$. Hence, this phase exists for,
\begin{equation}
0<x_w<1/2, \qquad \beta\leq 1/3
\end{equation}
Under these conditions, one can easily obtain that this phase exists only when $\mu>1$.
However, for $\mu\rightarrow\infty$, $x_w\rightarrow\infty$ that implies this phase vanishes in the limiting case of infinite number of particles and converges to HD-HD:HD-LD phase \cite{tian2020totally}.  
\subsubsection{\textit{HD-HD:HD-LD} Phase}
Without any loss of generality, in this phase, we assume both the segments of lane $L$ exhibit high density, whereas, left and right segment of $T$ display high and low density respectively. This phase exists when boundary controlling parameters satisfy the conditions presented in table \eqref{hh}.
\begin{table}[!htb]
\centering
\caption{\label{hh} Conditions for existence of asymmetric \textit{HD-HD:HD-LD} Phase}
\begin{tabular}{|c||c||c|}
\hline
& $L$ & $T$\\
\hline
\hline
Left segment & $\beta_{eff}<\min\{\alpha^{*},1/2\}$ & $\beta_{eff}<\min\{\alpha^{*},1/2\}$\\
Right segment & $\beta_{eff}<\min\{\alpha_{eff,L},1/2\}$ & $\alpha_{eff}<\min\{\beta,1/2\}$.\\
\hline
\end{tabular}
\end{table}\\
The density in each segment is given by,
\begin{eqnarray}
\rho_1^b=\rho_1^{k-1}=1-\beta_{eff} \quad && \text{and} \quad \rho_2^b=1-\beta,\\
\sigma_1^b=\sigma_1^{k-1}=1-\beta_{eff} \quad && \text{and} \quad \sigma_2^b=\sigma_2^{k+1}=\alpha_{eff,T}.
\end{eqnarray}
Since, the current is constant in each lane given in Eqn. \eqref{current} yields,
\begin{eqnarray}
\beta_{eff} &=& \beta,\\
\beta_{eff} &=& \alpha_{eff,T},
\end{eqnarray}
that implies,
\begin{equation}
\alpha_{eff,T} = \beta=\sigma^{k},
\end{equation}
giving density of particles from lane $T$ at intersected site.
The effective intrinsic rate is calculated utilising Eqn. \eqref{finite} that leads to,
\begin{eqnarray}
\alpha^* &=& \alpha\left(\dfrac{2\mu+2\beta-3}{2\mu}\right).
\end{eqnarray}
From Eqn. \eqref{ebeta} we have,
\begin{eqnarray}
\beta_{eff}=1-\rho^{k}-\beta,
\end{eqnarray}
that gives particle density of particles from lane $L$ as,
\begin{eqnarray}
\rho^{k} = 1-2\beta.
\end{eqnarray}
As a result the existence of conditions for this phase reduce to, 
\begin{equation}\label{conhlhl}
\dfrac{\alpha(3-2\mu)}{2(\alpha-\mu)}<\beta, \quad \beta\leq 1/3.
\end{equation}
Combing these conditions, we can easily conclude that this phase exists when $\mu>1.1$.
Here, in the limiting case $\mu\rightarrow\infty$, the conditions of existence reduces to $\alpha<\min\{\beta,1/3\}$, that for model with two intersected lanes and infinite number of particles \cite{tian2020totally}.

However, it is notable that not all combinations for asymmetric phases exist in the proposed model. For instance, suppose lane $L$ to exhibit high density in both the segments and lane $T$ to display low density in both the segments. The conditions of existence for this HD:HD-LD:LD phase are given by,
\begin{eqnarray}
\text{Lane $L$}\quad\beta<\{\alpha^*,1/2\},\\
\text{Lane $T$}\quad\alpha^*<\{\beta,1/2\}
\end{eqnarray}
which contradict each other. Similarly, other asymmetric phases cease to exist because either the conditions disagree or no values of $(\alpha,\beta)$ satisfy these conditions.

\begin{figure*}[!htb]
\centering
\subfigure[\label{25_08}]{\includegraphics[width = 0.4\textwidth,height=5.5cm]{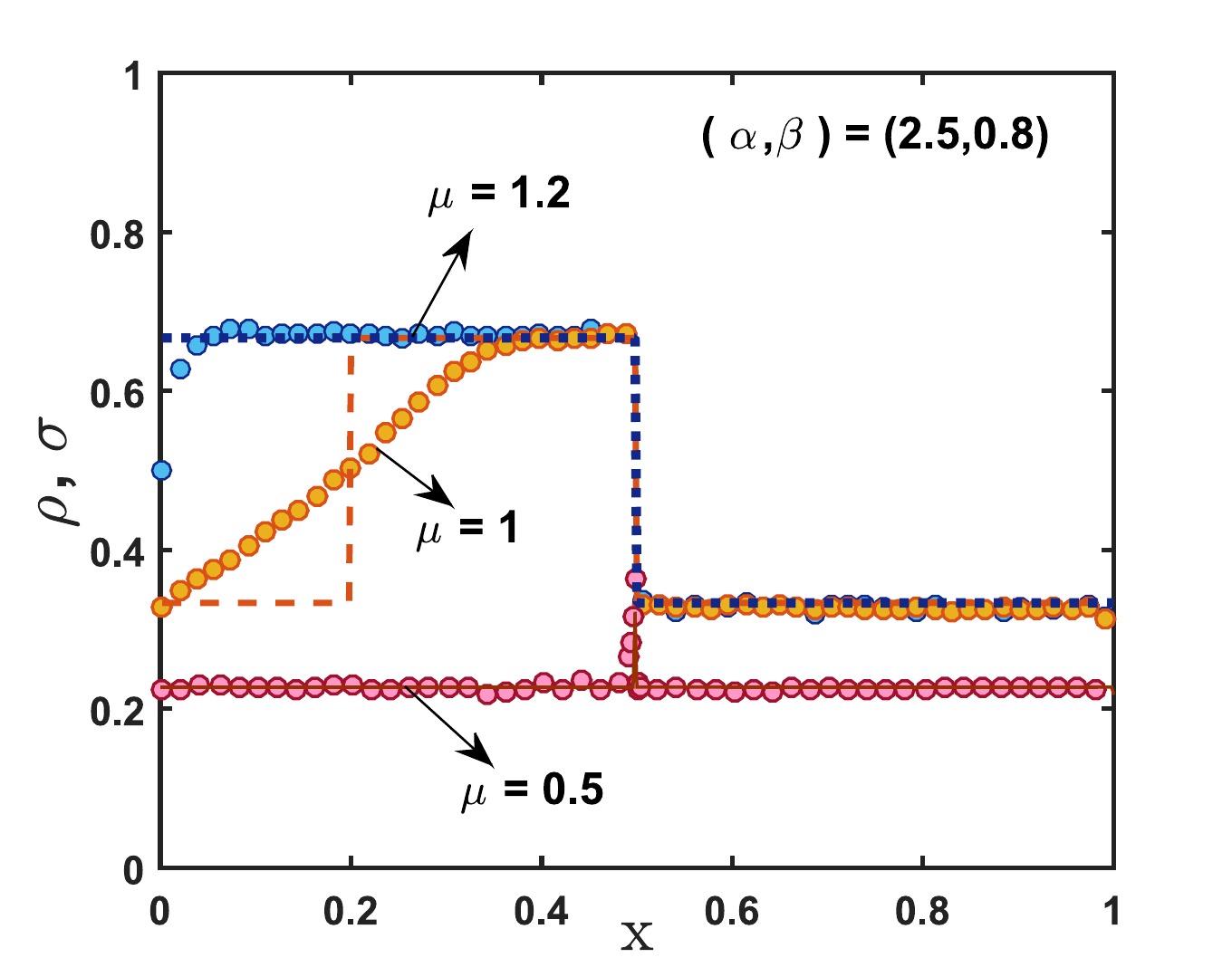}}
\subfigure[\label{2_02}]{\includegraphics[width = 0.4\textwidth,height=5.5cm]{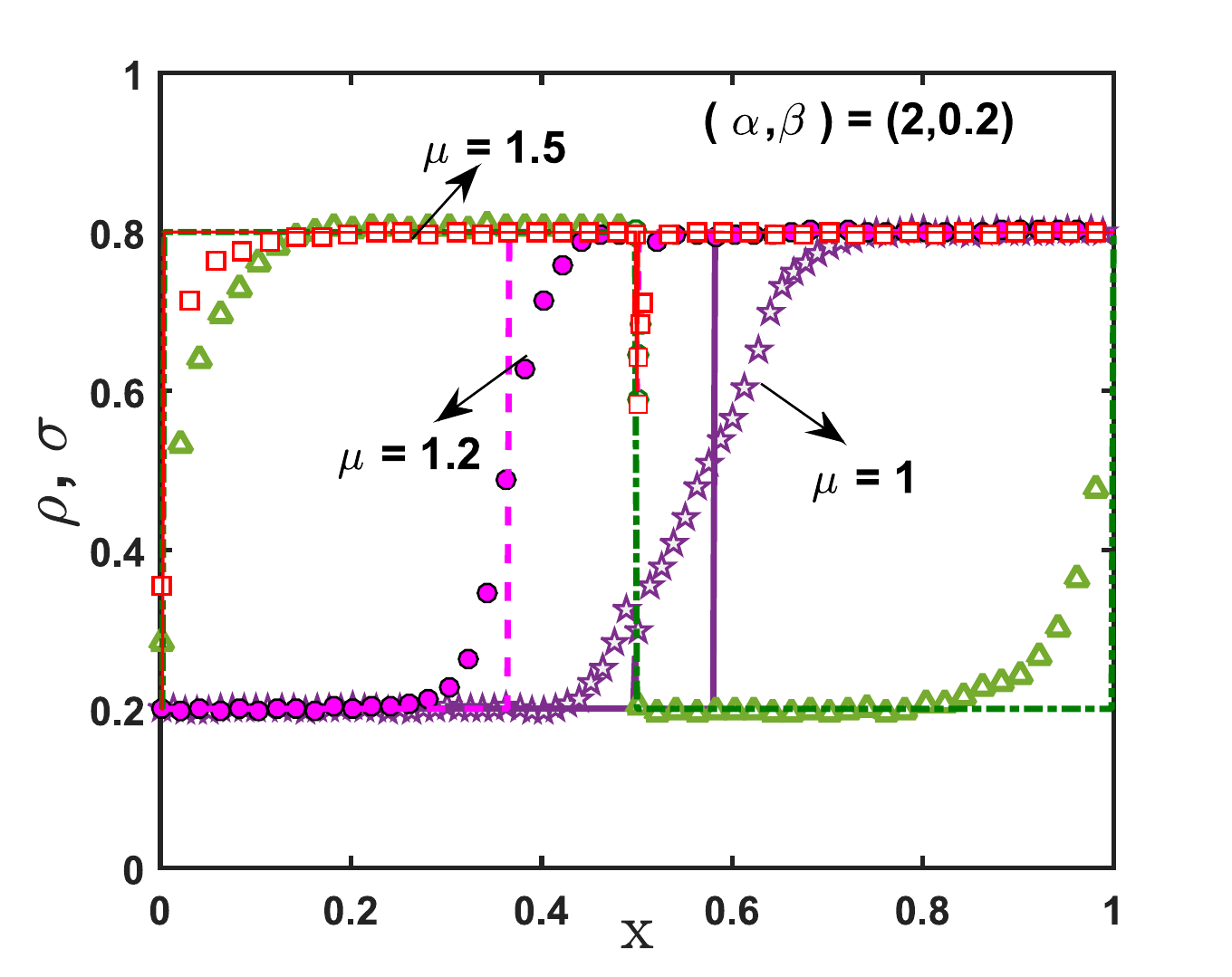}}
\caption{\label{effmu}Phase transitions with respect to $\mu$ for fixed boundary controlling parameters (a) $(\alpha,\beta)=(2.5,0.8)$ that presents transitions within symmetric phases LD-LD:LD-LD $\longrightarrow$ S-LD:S-LD $\longrightarrow$ HD-LD:HD-LD. (b) $(\alpha,\beta)=(2,0.2)$ illustrates phase transitions from symmetric to asymmetric phases LD-S:LD-S $\longrightarrow$ \textit{S-HD:HD-LD} $\longrightarrow$ \textit{HD-HD:HD-LD}. Since, for symmetric phase density profiles are same for both the lane, we plot for only one lane (say $L$). The solid lines represent theoretical results, while, markers denote the simulated outcomes.}
\end{figure*}

\section{Results and Discussions}
In this section, we exploit the results discussed in section \eqref{phases} to address the behaviour of system in ($\alpha,\beta$) plane depending on total number of particles in the system.
We aim to investigate the effect of finite resources in terms of $\mu=\frac{N_{tot}}{N}$ on the complex dynamical properties of the system. We observe qualitative as well as quantitative non-trivial effects on the topology of phase schema especially in terms of symmetry breaking and shock dynamics. To validate our theoretical outcomes we perform Monte Carlo simulations for system size $N=1000$. The computer simulations are carried out for $2 \times 10^9$ time steps and initial $5\%$ of the time steps are scraped to ensure the occurrence of steady state. The average densities in both the lanes are computed by considering time averages over an interval of $10N$. We observe that the theoretically computed density profiles, phase boundaries, shock positions match well with the simulations.
\subsection{Phase boundaries: Effect of $\mu$}
We have theoretically computed the existence of distinct stationary phases in terms of $\mu$ where we analysed that the symmetry of the system persists for $\mu\leq 1$. However, for $\mu>1$ the symmetry of the system is disrupted and asymmetric phases appears in the stationary phase diagram. So, to understand the effect of finite resources we elaborate two different cases and possible phase transitions originating in the system by monitoring the propagation of domain wall for varying boundary controlling parameters. 

\subsubsection{$\mu\leq1$}
When there are very less number of particles in the system i.e., $\mu \approx 0$, only one symmetric phase namely LD-LD:LD-LD appears in the entire phase regime as presented in Fig. \eqref{001} for $\mu=0.001$. This can be easily realised by a simple argument that due to scarcity of particles in the reservoir less number of particles are allowed to enter any of the lane, leading to low-density phase in each segment.  
Also, for lower values of $\beta$ the particles tend to accumulate at the right end of each lane. As a consequence, the boundary layer at the right boundary enters the bulk leading to the emergence of boundary induced shock in the right segment. Thus, a symmetric phase namely LD-S:LD-S phase appears in the phase diagram for $\mu>0$ as shown in Fig. \eqref{05} for $\mu=0.05$. The phase boundary between these two phases is obtained from Eqn. \eqref{slds} given by,
\begin{equation}
\alpha=\dfrac{\beta\mu}{\mu-2\beta},\quad\beta\leq 1/3.
\end{equation}

With significant increase in $\mu$ no qualitative changes are observed in the system except the shifting of phase boundaries resulting in the shrinkage of LD-LD:LD-LD phase and expansion of LD-S:LD-S phase. However, beyond a critical value $\mu_{C_1}$, due to interaction of particles at the intersected site a bulk induced shock emerges in the left segment of each lane resulting in a new symmetric regime S-LD:S-LD as shown in Fig. \eqref{1}. This critical value is computed from Eqn. \eqref{ssld} that yields $\mu_{C_1}=2/3$ and,
\begin{equation}
\alpha=\dfrac{\mu}{3\mu_{C_1}-2},\quad\beta\geq 1/3.
\end{equation}
However, for $\mu_{C_1}<\mu\leq\mu_{C_2}$ no new phase appears in the phase diagram, only quantitative changes are observed.
 
\subsubsection{$\mu>1$ (Symmetry Breaking)}
Distinctively, beyond a critical value $\mu_{C_2}$ in addition to the emergence of a new symmetric phase the system experiences rich topological changes with the occurrence of symmetry breaking phenomenon. As soon $\mu>\mu_{C_2}$, a new symmetric HD-LD:HD-LD phase emerges next to S-LD:S-LD phase resulting in the shrinkage of observed symmetric till now as shown in Fig. \eqref{12}. From Eqn. \eqref{muhlhl}, we observe that this phase exists for boundary controlling parameters satisfying,
\begin{equation}\label{bounsl}
\alpha=\dfrac{\mu_{C_2}}{3(\mu_{C_2}-1)},\quad\beta\geq 1/3.
\end{equation}
that yields the critical value $\mu_{C_2}=1$. In addition, due to interaction of particles at intersected site asymmetric \textit{S-HD:HD-LD} phase emerges in $(\alpha,\beta)$ plane for which the phase boundary is obtained from Eqn. \eqref{sshd} that leads to,
\begin{equation}
\alpha=\dfrac{\beta\mu_{C_2}}{\mu_{C_2}-1},\quad\beta\leq 1/3.
\end{equation}
This is because with increase in number of particles, the boundary induced shock in LD-S:LD-S phase absorbs the incoming particles and travel towards the left side of each lane. As soon it reaches the intersected site, the particle at intersected site totally blocks the flow of particles resulting in HD-LD phase in one lane (say lane $L$). For other lane ($T$) the boundary induced shock crosses the intersected site and shifts to the left segment.
\begin{figure*}[!htb]
\centering
\subfigure[\label{spls}$\beta=0.2$]{\includegraphics[width = 0.35\textwidth,height=5.5cm]{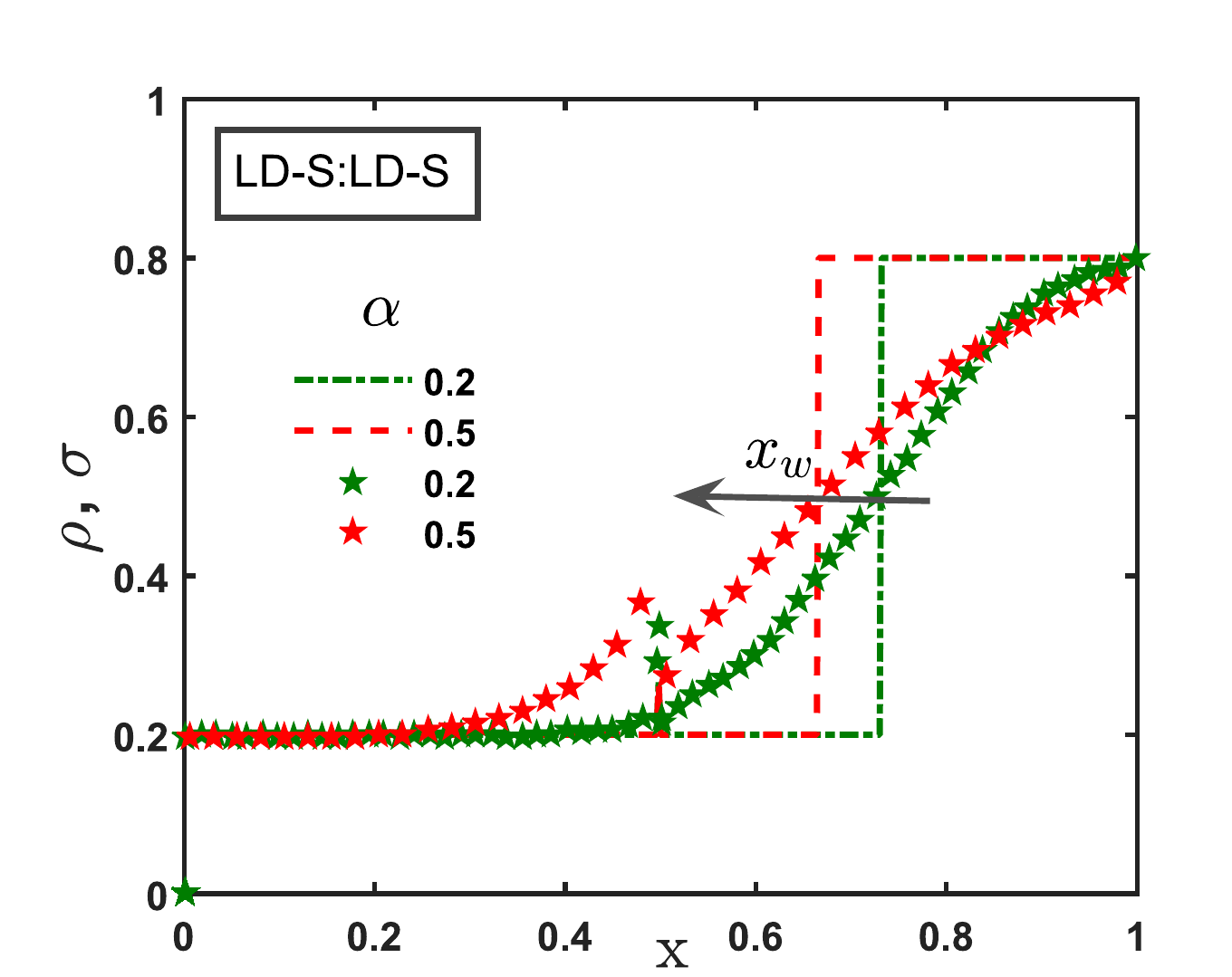}}
\subfigure[\label{spsl}$\beta=0.8$]{\includegraphics[width = 0.35\textwidth]{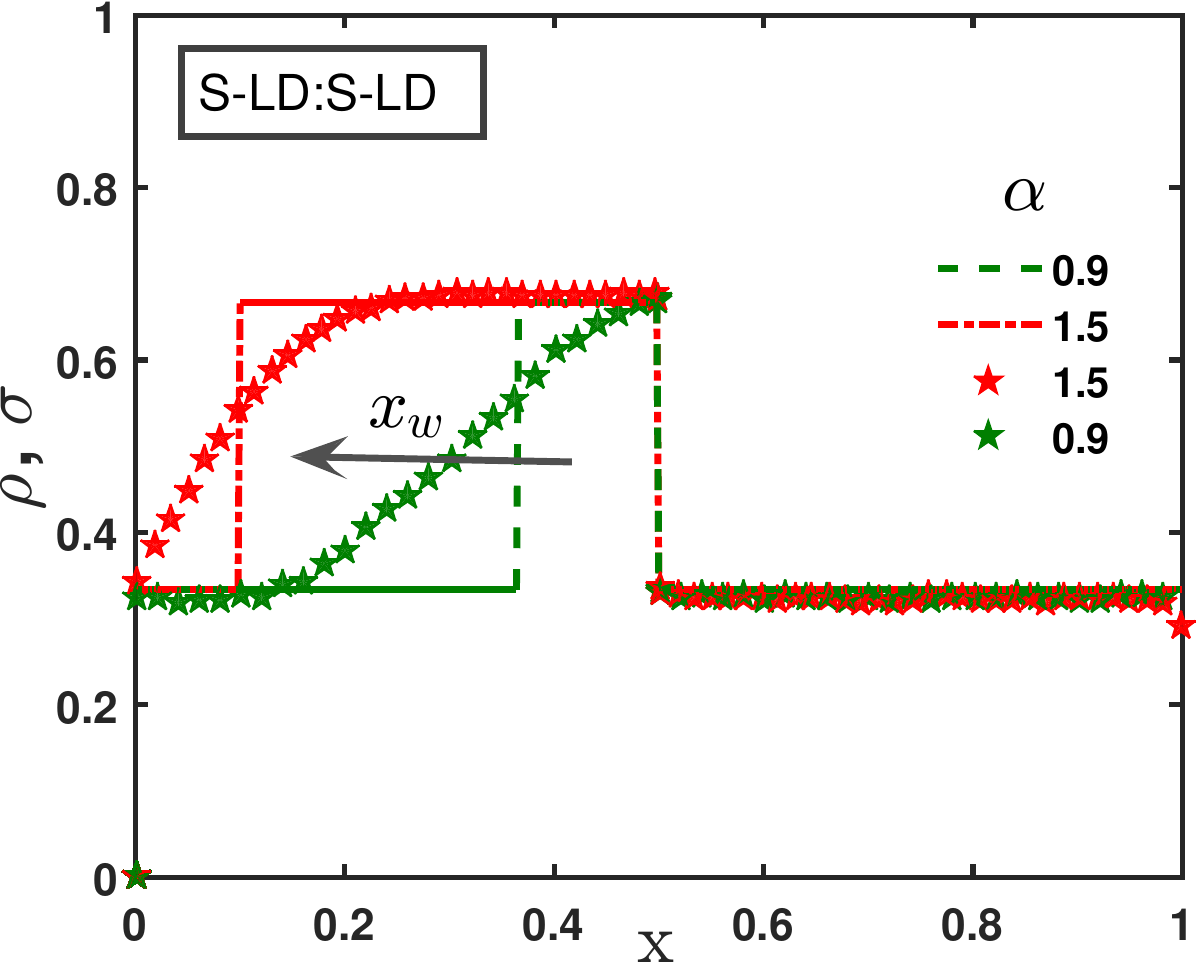}}
\caption{\label{spsym}Symmetric Phases: Movement of shock with increasing values of $\alpha$ for a chosen $\mu=1.2$ in (a) LD-S:LD-S phase and (b) S-LD:S-LD for $\beta=0.2$ and $0.8$ respectively. As already discussed, the symmetric phase satisfies Eqn. \eqref{symcon1} and \eqref{symcon2}. Therefore, without any loss of generality, we plot the density profile of particles of only one lane (say $L$). The same results are pertinent for the other lane $T$. The solid, dashed lines represent the theoretical computed results. While, markers denote the simulated results.}
\end{figure*}
\begin{figure*}[!htb]
\centering
\includegraphics[width = 0.35\textwidth]{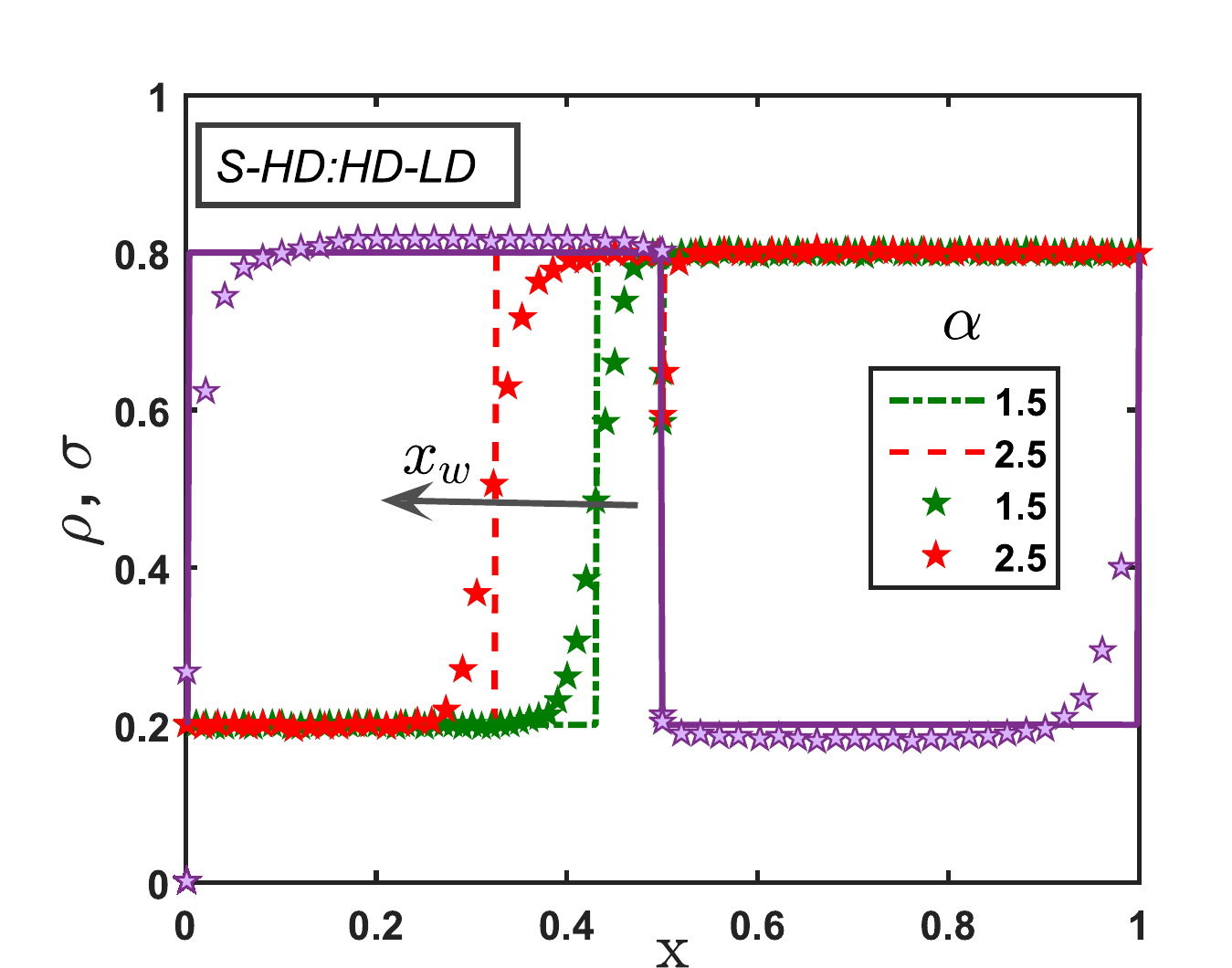}
\caption{\label{spsh}Asymmetric Phase: Movement of shock with increasing values of $\alpha$ for $\mu=1.2$ and $\beta=0.2$ in \textit{S-HD:HD-LD} phase. One of the lane portrays HD-LD phase for chosen parameters marked with purple markers and solid line. However, the other lane manifests S-HD phase for which the shock position varies with $\alpha$. As $\alpha$ increases the domain wall sweeps to the left of the lane. The solid, dashed lines represent the theoretical computed results and, markers denote the simulated results.}
\end{figure*} On further increasing $\mu$ after a crucial value $\mu_{C_3}$ an additional asymmetric phase \textit{HD-HD:HD-LD} emerges in the phase schema (as prescribed in Fig. \eqref{15}) for,
\begin{equation}
\alpha=\dfrac{2\beta\mu_{C_3}}{2(\mu_{C_3}+\beta)-3},\quad\beta\leq 1/3,
\end{equation}
and crucial value is obtained as $\mu_{C_3}=1.1$. This happens because as soon a particle enters the lane for entry rate satisfying Eqn. \eqref{conhlhl}, it is quickly assimilated by the segment that exhibits shock in \textit{S-HD:HD-LD} phase. Beyond this crucial value $\mu_{C_3}$, we observe only quantitative alterations in the phase diagram as presented in Fig. \eqref{2} for $\mu=2$. Even for finite value of $\mu$ the phase diagram of proposed model converges to that of two intersected TASEP model in ref.\cite{tian2020totally}. However, in the limiting case $\mu\rightarrow \infty$, it is clearly evident from Eqns. \eqref{ssld}, \eqref{slds} and \eqref{sshd} that the boundary and bulk induced shock disappears, and we retrieve phase diagram for two intersected lanes with infinite number of particles.\\

\subsection{Density Profiles and Phase Transitions}
The density profiles attributed to symmetric and asymmetric stationary phases are presented in Fig. \eqref{densym} and \eqref{denasym} respectively are found to agree well with Monte Carlo simulations. However, the shock present in Fig. \eqref{lsls} and \eqref{slsl} becomes sharper for increasing values of $N$. Due to intersection of lanes, the density profile of LD-S:LD-S, LD-LD:LD-LD, \textit{HD-HD:HD-LD} and \textit{S-HD:HD-LD} phase admits a kink at the intersected site as clearly visible in Fig. \eqref{lsls}, \eqref{ldld}, \eqref{hdhdhdld} and \eqref{shd} respectively.\\
\indent To further inspect a deeper insight in the phenomenon of SSB, we probe particle density histograms $P(\rho_1,\rho_2)$ and $P(\sigma_1,\sigma_2)$, where $\rho_j$ and $\sigma_j$ are instantaneous particle densities on segment $j$ ($j=1,2$). For $\alpha=2.5$ and $\beta=0.8$, we present typical density histogram for asymmetric phases \textit{S-HD:HD-LD} and \textit{HD-HD:HD-LD} with $\mu=1.2$ and $\mu=1.5$ respectively. One can clearly see in Fig. \eqref{hishl} that the peaks in distributions are achieved for $\rho_1=\rho_2>1/2$ and $\sigma_1>1/2$, $\sigma_2<1/2$ that corresponds to \textit{HD-HD:HD-LD} phase. Fig. \eqref{hisshd} demonstrates that the peak occurs for $0<\rho_1<1$, $\rho_2>1/2$ and $\sigma_1>1/2$, $\sigma_2<1/2$ portraying \textit{S-HD:HD-LD} phase.\\

In order to visualise phase transitions with respect to $\mu$ we chose particular values of $\alpha,\beta$ and plot Fig. \eqref{effmu}. For chosen parameters $\alpha=2.5$ and $\beta=0.8$, in Fig. \eqref{25_08} we portray transitions within symmetric phases LD-LD:LD-LD $\longrightarrow$ S-LD:S-LD $\longrightarrow$ HD-LD:HD-LD. When $\mu=0.5$ the particles exhibit symmetric LD-LD:LD-LD phase with a kink in the density profile at intersected site. With increase in $\mu=1.2$, the density at intersected site increases as also evident from Eqn. \eqref{int}, that indicates the existence of symmetric S-LD:S-LD phase. Further increasing $\mu=1.5$, the bulk induced shock transforms into HD regime and leads to the occurrence of symmetric HD-LD:HD-LD phase. Similarly, Fig. \eqref{2_02} illustrates the phase transitions from symmetric to asymmetric phases LD-S:LD-S $\longrightarrow$ \textit{S-HD:HD-LD} $\longrightarrow$ \textit{HD-HD:HD-LD} for $\alpha=2$ and $\beta=0.2$. When $\mu=1$ particles manifest boundary induced shock in right segment of both the lanes displaying symmetric LD-S:LD-S phase. Increasing $\mu=1.2$, due to availability of ample number of particles this shock stabilises and transforms into asymmetric \textit{S-HD:HD-LD} phase which further converges to asymmetric \textit{HD-HD:HD-LD} phase for $\mu=1.5$.

To summarize, there exist maximum six stationary phases in the overall system including four symmetric and two asymmetric phases. The complexity of phase diagram shows non-monotonic behaviour with increasing values of $\mu$. Initially, there exists only one phase while in the midrange dynamics becomes complex and six phases are observed. Besides, the intuitive observations of the effect of finite resources, the appearance and disappearance of phases is examined from theoretically computed phase boundaries. In the limiting case $\mu\rightarrow \infty$, we have explained in section \eqref{phases}, how effective intrinsic rate $\alpha^*$ approaches to $\alpha$ in each phase. Consequently, topological structure of phase schema is modified and the number of phases drastically reduces from six to three \cite{tian2020totally}.

\subsection{Shock Dynamics}\label{Shock}
In the above sections, we have seen that due to finite number of particles in the system, two types of shock emerges in the dynamical regimes. It has been observed that a bulk induced shock exists in symmetric S-LD:S-LD phase. Whereas, a boundary induced shock persists in a symmetric LD-S:LD-S as well as a asymmetric \textit{S-HD:HD-LD} phase. The shock emerging in the right segment of any lane is boundary induced, while, that appearing in left segment might be boundary or bulk induced. Here, to discuss the phase transitions arising due to propagation of shock in ($\alpha-\beta$) plane we fixed parameter and vary boundary controlling parameters $(\alpha,\beta)$.

Beginning with the boundary induced shock in symmetric LD-S:LD-S phase appearing for $\beta<1/3$, the shock lies in the right segment of both the lanes determined by Eqn. \eqref{slds}. For $\beta=0.2$, we plot Fig. \eqref{spls} where one can notice that as $\alpha$ increases shock propagates toward the left side of right segment. This means, if more number of particles are allowed to enter the lattice for a significant choice of $\mu$, the right segment incorporates these particles tending towards a high dense region. As soon this wall reaches the intersected site, due to interactions other lane is forced to exhibit low density of particles. Thus, next to this phase, with increasing $\alpha$, a asymmetric phase \textit{S-HD:HD-LD} appears in the steady-state phase diagram. For this phase, the position of boundary induced shock lies within the range $[0,1/2]$ and can be computed using Eqn. \eqref{sshd}. With increase in $\alpha$, the particles are absorbed by the segment specifying S phase. As a consequence, shock moves toward the left of the lane as shown in Fig. \eqref{spsh} for $\beta=0.2$ and beyond a critical value of $\alpha$, the left lane exhibits HD indicating the appearance of asymmetric \textit{HD-HD:HD-LD} phase. 

For $\beta>1/3$, a bulk induced shock in symmetric S-LD:S-LD phase appears in the phase schema. In this phase the particles in left segment of both the lanes portray a discontinuity. The explicit expression for the location of this wall is computed in Eqn. \eqref{ssld} that suggests for a fixed $\beta$, as more number of particles are permitted to enter, the shock in both the lanes sweeps to the left of the lane. This can be easily noticed from Fig. \eqref{spsl} where we can clearly see that as $\alpha$ increases for $\beta=0.8$, the HD part of S increases and after a crucial value of $\alpha$ given in Eqn. \eqref{bounsl} shock vanishes. As a result, left segments of both the lanes attain HD phase leading to the occurrence of symmetric HD-LD:HD-LD phase.

\section{Conclusion}
In this work, we have studied a specific variant of network TASEP model with two intersected lanes, a class of minimal models for transportation phenomenon. The two extreme ends of each lane are connected to a reservoir having finite number of particles. The intersection of lanes introduces an inhomogeneity in the system that is suitably dealt by considering effective entry and exit rates. Even though, the particles interact at intersected site, mean-field approximation works well to theoretically investigate various crucial steady-state properties of the system such as density profiles, phase transitions and phase boundaries. The theoretical predictions are validated through extensive Monte Carlo simulations.

We extensively probe the effect of finite resources on the phenomenon of spontaneous symmetry breaking, since the same persists for infinite number of particles. With an increase in number of particles, crucial qualitative and quantitative changes in the topology of phase diagram are observed. The symmetry of phase schema is preserved till the total number of particles do not exceed the total number of sites. However, as soon as more number of particles are available than the number of sites, the symmetry of the phase diagram is disrupted. The interaction of particles at intersected site is responsible for the symmetry breaking phenomenon. There exist maximum six possible stationary regimes in the system including four symmetric and two asymmetric phases. The exact number and location of phases depend on the number of particles in the system. The existence of asymmetric phases is explored by delineating density histograms using computer simulations. In addition, a symmetric phase exhibits a bulk and boundary induced shock in each lane. Whereas, a asymmetric phase manifests a boundary induced shock in one of the lanes due to interactions of particles at intersected site. The density in the shock profile is estimated by employing domain-wall approach. We explicitly calculate the phase boundaries to determine dynamic regimes and location of different phases. Also, by monitoring the movement of shock we describe the phase transitions as more particles are allowed to enter.

The proposed study can be extended to analyse more complex dynamics in the network of intersected lanes when inter-lane switching of particles is allowed in the bulk as well as while jumping from the intersected site. Also, several generalisations of the model can be explored in future by considering various other processes such as the interplay with the non-conserving dynamics, extended particle size, particle-particle interactions etc.. 

\section*{Acknowledgement}
AK Gupta acknowledges support from DST-SERB, Govt. of India (Grant CRG/2019/004669).
%

\end{document}